\documentclass[fleqn,usenatbib]{mnras}

\usepackage{newtxtext,newtxmath}
\usepackage[T1]{fontenc}
\usepackage{ae,aecompl}

\DeclareRobustCommand{\VAN}[3]{#2}
\let\VANthebibliography\thebibliography
\def\thebibliography{\DeclareRobustCommand{\VAN}[3]{##3}\VANthebibliography}

\usepackage{graphicx}	% Including figure files
\usepackage{amsmath}	% Advanced maths commands
\usepackage{subcaption}
\usepackage{color}
\usepackage{ulem}
\usepackage{hyperref}
\usepackage{soul}
\newcommand{\grale}{\textsc{grale }}
\newcommand{\lenstool}{\textsc{lenstool }}
\newcommand{\hst}{{\it HST}}
\title[{\sc grale} inversion of A370 using BUFFALO data]{Further support for a trio of mass-to-light deviations in Abell 370: free-form {\sc grale} lens inversion using BUFFALO strong lensing data}

\author[A. Ghosh et al.]{Agniva Ghosh,$^{1}$\thanks{E-mail: ghosh116@umn.edu (AG); llrw@umn.edu (LLRW)} 
Liliya L.R. Williams,$^{1}$\footnotemark[1], 
Jori Liesenborgs,$^{2}$ 
Ana Acebron,$^{3}$ 
\newauthor 
Mathilde Jauzac,$^{4,5,6,7}$
Anton M. Koekemoer,$^{8}$ 
Guillaume Mahler, $^{4,5}$
Anna Niemiec,$^{4,5}$
\newauthor
Charles Steinhardt,$^{9,10}$
Andreas L. Faisst,$^{11}$
David Lagattuta$^{4,5}$ 
 and 
\newauthor
Priyamvada Natarajan$^{12,13}$
\\
$^{1}$School of Physics and Astronomy, University of Minnesota, 116 Church Street SE, Minneapolis, MN 55455, USA\\
$^{2}$UHasselt -- tUL, Expertisecentrum voor Digitale Media, Wetenschapspark 2, B-3590, Diepenbeek, Belgium\\
$^{3}$Dipartimento di Fisica, Università degli Studi di Milano, via Celoria 16, I-20133 Milano, Italy\\
$^{4}$Centre for Extragalactic Astronomy, Durham University, South Road, Durham DH1 3LE, UK\\
$^{5}$Institute for Computational Cosmology, Durham University, South Road, Durham DH1 3LE, UK\\
$^{6}$Astrophysics Research Centre, University of KwaZulu-Natal, Westville Campus, Durban 4041, South Africa\\
$^{7}$School of Mathematics, Statistics \& Computer Science, University of KwaZulu-Natal, Westville Campus, Durban 4041, South Africa\\
$^{8}$Space Telescope Science Institute, 3700 San Martin Drive, Baltimore, MD 21218, USA\\
$^{9}$Cosmic Dawn Center, Denmark\\
$^{10}$Niels Bohr Institute, University of Copenhagen, Lyngbyvej 2, DK-2100 Copenhagen Ø, Denmark\\
$^{11}$Caltech/IPAC, MS 314-6, 1200 E. California Blvd. Pasadena, CA 91125, USA\\
$^{12}$ Department of Astronomy, Yale University, 52 Hillhouse Avenue, New Haven, CT 06511, USA \\
$^{13}$ Department of Physics, Yale University, P.O. Box 208121, New Haven, CT 06520, USA \\
}

\date{Accepted XXX. Received YYY; in original form ZZZ}

\pubyear{2021}

\begin{document}

\label{firstpage}
\pagerange{\pageref{firstpage}--\pageref{lastpage}}
\maketitle

\begin{abstract}
We use the Beyond Ultra-deep Frontier Fields and Legacy Observations (BUFFALO) strong lensing image catalog of the merging galaxy cluster Abell 370 to obtain a mass model using the free-form lens inversion algorithm \textsc{grale}. The improvement of the strong lensing data quality results in a lens plane rms of only 0.45 arcsec, about a factor of two lower than that of our existing HFF v4 reconstruction. 
% {\sout{Since the number of images is nearly the same in both, we attribute the improvement to the improvement in the data quality.}}
{ We attribute the improvement to spectroscopic data and use of the full reprocessed \hst mosaics.}
In our reconstructed mass model, we found indications of three distinct mass features in Abell 370: (i) a $\sim\!35$ kpc offset between the northern BCG and the nearest mass peak, (ii) a $\sim\!100$ kpc mass concentration of roughly critical density $\sim\!250$  kpc east of the main cluster, and (iii) a probable filament-like structure passing N-S through the cluster. While (i) is present in some form in most publicly available reconstructions spanning the range of modeling techniques: parametric, hybrid, and free-form, (ii) and (iii) are recovered by only about half of the reconstructions. We tested our hypothesis on the presence of the filament-like structure by creating a synthetic cluster - Irtysh IIIc - mocking the situation of a cluster with external mass. We also computed the source plane magnification distributions. Using them we estimated the probabilities of magnifications in the source plane, and scrutinized their redshift dependence. Finally, we explored the lensing effects of Abell 370 on the luminosity functions of sources at $z_s=9.0$, finding it consistent with published results.

\end{abstract}

\begin{keywords}
galaxies: clusters: individual: Abell 370 -- gravitational lensing: strong
\end{keywords}

%%%%%%%%%%%%%%%%%%%%%%%%%%%%%%%%%%%%%%%%%%%%%%%%%%

%%%%%%%%%%%%%%%%% BODY OF PAPER %%%%%%%%%%%%%%%%%%

\section{Introduction}
For well over a decade lens modelling of clusters of galaxies has been an extraordinarily useful tool for revealing intrinsic properties of lensed sources at high redshifts. Clusters of galaxies can magnify faint distant background sources by increasing their angular extent and observed fluxes, and thus pushing them above the detection threshold of contemporary telescopes \citep{Schneider1992,Bartelmann2010}. By calculating the volume number densities of these sources one can estimate the ultraviolet luminosity function, which helps in the understanding of the early galaxy evolution, their star formation rate, and thus their role in the reionization of the universe \citep{Bouwens2017,Livermore2017,Ishigaki2018,Atek2018}.

To utilise clusters of galaxies as natural telescopes one needs to characterize their uneven optics, i.e. obtain magnification maps, which are directly related to the mass distributions in the clusters \citep{Kneib2011}. This characterization is achieved by strong lens reconstruction algorithms which use multiple images of background sources. Projected mass models of clusters obtained by strong lensing inversion techniques give the most detailed maps of cluster dark matter distribution obtained by any method.

With the contemporary observational techniques, the strongest cluster lenses have of the order of $\sim100$ multiple images spread across $\lesssim 1$ arcmin. This number of constraints is inadequate to break all lensing degeneracies, and thus the reconstructed mass distributions are usually not unique \citep{Limousin2016,Priewe2017}. However, since the inception of the {\it Hubble} Frontier Fields survey (HFF; PI: J. Lotz) using the {\it Hubble Space Telescope} (\hst), a range of different lens inversion methods - which use different modelling assumptions and procedures - have made it possible to estimate systematic uncertainties based on different reconstructions of the same clusters of galaxies \citep{Lotz2017}. Comprehensive comparison projects were carried out based on synthetic clusters \citep{Meneghetti2017}, as well as observed HFF clusters \citep{Priewe2017,Gonzalez2018}, exploring systematic uncertainties in mass reconstructions \citep{Raney2020a},  probability of magnifications \citep{Vega2019}, and recovered luminosity function parameters of high-$z$ galaxies \citep{Bouwens2017a,Livermore2017,Ishigaki2018}.

More recently, the Beyond Ultra-deep Frontier Fields and Legacy Observations (BUFFALO; PI: C. Steinhardt) program has embellished the HFF galaxy clusters by extending fields that are already covered by multiwavelength data from the {\it Spitzer} Space Telescope, the {\it Chandra} and the {\it XMM-Newton} X-ray observatories and other ground based observatories \citep{Steinhardt2020}. It has observed four times more than the existing HFF sky area for all six HFF clusters in five \hst~filters. Additional spectroscopic data from MUSE \citep{Lagattuta2017,Lagattuta2019,Richard2021} has also significantly contributed towards better lens reconstructions.

Reconstruction models can be broadly classified into two categories: parametric and free-form. Parametric models, which are more widely used, assign simple mass profiles - for example, Navarro-Frenk-White \citep[NFW;][]{Navarro1996} profiles, Pseudo Isothermal Elliptical Mass Distributions \citep[PIEMD; ][]{Kovner1993}, pseudo-Jaffe distributions \citep[PJ;][]{Keeton2001} - to observed galaxies, in association with luminosity-mass scaling relations, like the relation of \cite{Faber1976}. In other words, the mass substructure in parametric models is always closely tethered to the observed luminosity distribution in clusters. 
The cluster-scale dark matter distribution is represented by the same or similar simple profiles, but with larger scale lengths, or by smoothed out galaxy light distributions \citep{Natarajan1997,Zitrin2009}.

In contrast to parametric models, free-form models (such as our method \grale\!, see Section~\ref{sec:grale}) do not assume any relation between the distribution of mass and light on the sky. In fact, \grale does not have enough spatial resolution to account for the most compact baryonic distributions of many cluster galaxies \citep[see][]{Liesenborgs2020}. The advantage of free-form methods is that they use the lensed images only, making them very sensitive to the information content of the lensed images. Parametric methods include strong priors in addition to images, so reconstructions are a compromise between image constraints and priors. While these priors are well motivated by astrophysics, they describe the average properties of galaxies and clusters, not specific properties that may differ between clusters and between individual galaxies, or deviate from averages, especially in merging clusters.  Because of their unmitigated sensitivity to lensed images, free-form methods can detect cluster mass features that elude parametric methods. This is possible even with smaller numbers of images than in HFF, or with less than perfect data (see Section~\ref{sec:recres}).

Abell 370 (hereafter A370), the subject of this paper, is the first massive cluster observed to be hosting gravitationally lensed images of background sources, namely the giant luminous arc in the southern part of the cluster. The cluster was studied in depth and modelled by various groups starting in the mid 1980s \citep{Soucail1987,Soucail1988,Hammer1987,Hammer1989,Lynds1989,Kovner1989,Kneib1993}. The first non-parametric reconstruction was done by \citet{Abdelsalam1998}. One of the first strong lensing models using the multi-color images from the refurbished \hst~Advanced Camera for Surveys (ACS) was created by \citet{Richard2010}, and weak lensing analyses followed \citep[e.g.,][]{Medezinski2011,Umetsu2011}. Since the inclusion of A370 as one of the six HFF clusters in 2013, deeper imaging data became available. Eight different modelling groups produced mass distributions of the central cluster region, which are publicly available from the Mikulski Archive for Space Telescopes (MAST) website\footnote{\url{https://archive.stsci.edu/pub/hlsp/frontier/abell370/models/}}.

Several parametric codes were used for A370 modelling in the HFF project. \textsc{lenstool} was used by the Clusters As TelescopeS \citep[CATS;][]{Richard2014}, the Johnson-Sharon groups \citep{Johnson2014} and the works of \citet{Lagattuta2017,Lagattuta2019}. It utilizes a Bayesian Markov Chain Monte-Carlo sampler to optimize the modelling parameters. The overall mass distribution is modelled as a superposition of smooth large-scale potentials such as PIEMD and small-scale substructures that are associated with the locations of cluster member galaxies \citep{Natarajan1997}. \citet{Kawamata2018} used \textsc{glafic}, a parametric method which models the halo components by an NFW profile and cluster member galaxies by a pseudo-Jaffe (PJ) profile. The Keeton group used the parametric \textsc{lensmodel} code, which in addition to large-scale halos for the dark matter and/or hot gas, and small-scale halos for cluster members, also includes small-scale halos to model line-of-sight (LoS) galaxies \citep{Raney2020}. Among other parametric methods which submitted A370 models using HFF data but do not have a specific publication with their models, are Light Traces Mass \citep[LTM;][]{Zitrin2016}, and PIEMDeNFW \citep{Zitrin2013} used by the Zitrin group. LTM assumes that the dark matter distribution in the cluster can be approximately represented by the smoothed out luminous distribution.  PIEMDeNFW uses PIEMD profiles for the cluster members, and elliptical NFW or PIEMD profile for the dark matter distributions.

Among the non-parametric modelers, Bradac-Hoag group used the free-form code \textsc{swunited} \citep{Strait2018} - the only A370 model which combines strong and weak lensing data. It uses an iterative $\chi^2$ minimization process to solve for the gravitational potential on a grid. Our group worked with \grale to model A370 with pre-HFF \citep{Mohammed2016} and HFF data. In addition to the above reconstructions, the Diego group used the hybrid method \textsc{wslap+} \citep{Diego2018b, Vega2019}, which decomposes the mass distribution into a free-form grid component for the diffuse mass, and a parametric component for compact member galaxies. 

The main motivation behind this work is to present a comprehensive analysis of the cluster A370 using our free-form lens inversion method \textsc{grale} and utilizing the new strong lensing data from the BUFFALO collaboration, for the first time.
The structure of this paper is as follows: Section~\ref{sec:grale} describes our free-form reconstruction method, \grale\!, and the input images from the A370 BUFFALO catalog.  We discuss the reconstructed mass distribution in Section~\ref{sec:recres}. In that section, we concentrate on how well mass follows light in the very core of the cluster (\ref{sec:bcg}), and on two possible mass features: a $\sim 100$  kpc mass clump of roughly critical density in the eastern part of the cluster (\ref{sec:substructure}), and a filament-like structure passing N-S through the cluster (\ref{sec:filament}). 
In Section~\ref{sec:mag}, we map the reconstructed magnification distributions in the source plane. Using these, we estimate the probabilities of magnification in the source plane (\ref{sec:magprob}), and explore the redshift dependence of these probabilities as an indication of mass substructure (\ref{sec:magz}). Finally, in Section~\ref{sec:LF}, we estimated the lensed luminosity function and its uncertainties by convolving the reconstructed magnification distribution with the classical Schechter luminosity function with the best-fit parameter values taken from the literature.

Throughout this paper we use the $\Lambda$CDM model of cosmology: flat, with matter density, $\Omega_m = 0.3$, cosmological constant density, $\Omega_{\Lambda}= 0.7$, and the dimensionless Hubble constant $h = 0.7$. The redshift of A370 is 0.375.  The center of the reconstruction region is at R.A.$=39.970^{\circ}$, Decl.$=-1.577^{\circ}$. At the redshift of the cluster, 1 arcsec corresponds to 5.15  kpc.

\section{Reconstruction Method: Grale}
\label{sec:grale}
\subsection{Method}

The lens inversion method used in this paper is based on the reconstruction code \grale\!\footnote{\url{https://research.edm.uhasselt.be/jori/grale2/}}. The publicly available \grale software implements a flexible, free-form, adaptive grid lens inversion method, based on a genetic algorithm \citep{Liesenborgs2006a,Liesenborgs2007,Mohammed2014,Meneghetti2017}. It is ideally suited for reconstructions with numerous multiple images, available with \hst~data.  The fact that the number of its model parameters exceeds the number of data constraints allows a fuller exploration of degenerate mass distributions
\citep{sebesta2016,sebesta2019,williams2018,Mohammed2014,williams2019}. We refer the readers to see section 3.1 of \citet[hereafter, G20]{Ghosh2020}, for a concise description of the modus operandi of \grale\!. {In this work we used single lens plane inversion with \grale. The current stable version of \grale is not capable of doing multi-lens plane reconstruction.}

\subsection{Input}
\begin{table}
\centering
    \caption{Summary of the strong lensing images used from the BUFFALO catalog of A370. Full table is available as online supplementary material. [L17: \citet{Lagattuta2017}, K18: \citet{Kawamata2018}, L19: \citet{Lagattuta2019}]}  
    \label{tab:imgdata}
    \begin{center}
\begin{tabular}{lllllr}
\hline
ID & R.A       & Dec        & $z_{\rm used}$    & Reference & Quality         \\      
    & (deg)     & (deg)      &             &           &                 \\
\hline 
1.1      & 39.967047 & -1.5769172 & 0.8041            & L17, L19   & \textsc{gold}   \\
1.2      & 39.976273 & -1.5760558 & 0.8041            & L17, L19   & \textsc{gold}   \\
1.3      & 39.968691 & -1.5766113 & 0.8041            & L17, L19   & \textsc{gold}   \\
\hline 
2.1      & 39.973825 & -1.584229  & 0.7251            & L17, L19   & \textsc{gold}   \\
2.2      & 39.971003 & -1.5850422 & 0.7251            & L17, L19   & \textsc{gold}   \\
2.3      & 39.968722 & -1.5845058 & 0.7251            & L17, L19   & \textsc{gold}   \\
2.4      & 39.969394 & -1.5847328 & 0.7251            & L17, L19   & \textsc{gold}   \\
2.5      & 39.96963  & -1.5848508 & 0.7251            & L17, L19   & \textsc{gold}   \\
\hline
3.1      & 39.965658 & -1.566856  & 1.9553            & L17, L19   & \textsc{gold}   \\
3.2      & 39.968526 & -1.5657906 & 1.9553            & L17, L19   & \textsc{gold}   \\
3.3      & 39.978925 & -1.5674624 & 1.9553            & L17, L19   & \textsc{gold}   \\
\hline
$\cdots$ \\
\hline
\end{tabular}
\end{center}
\end{table}

\begin{figure*}
	\includegraphics[width=0.8\textwidth]{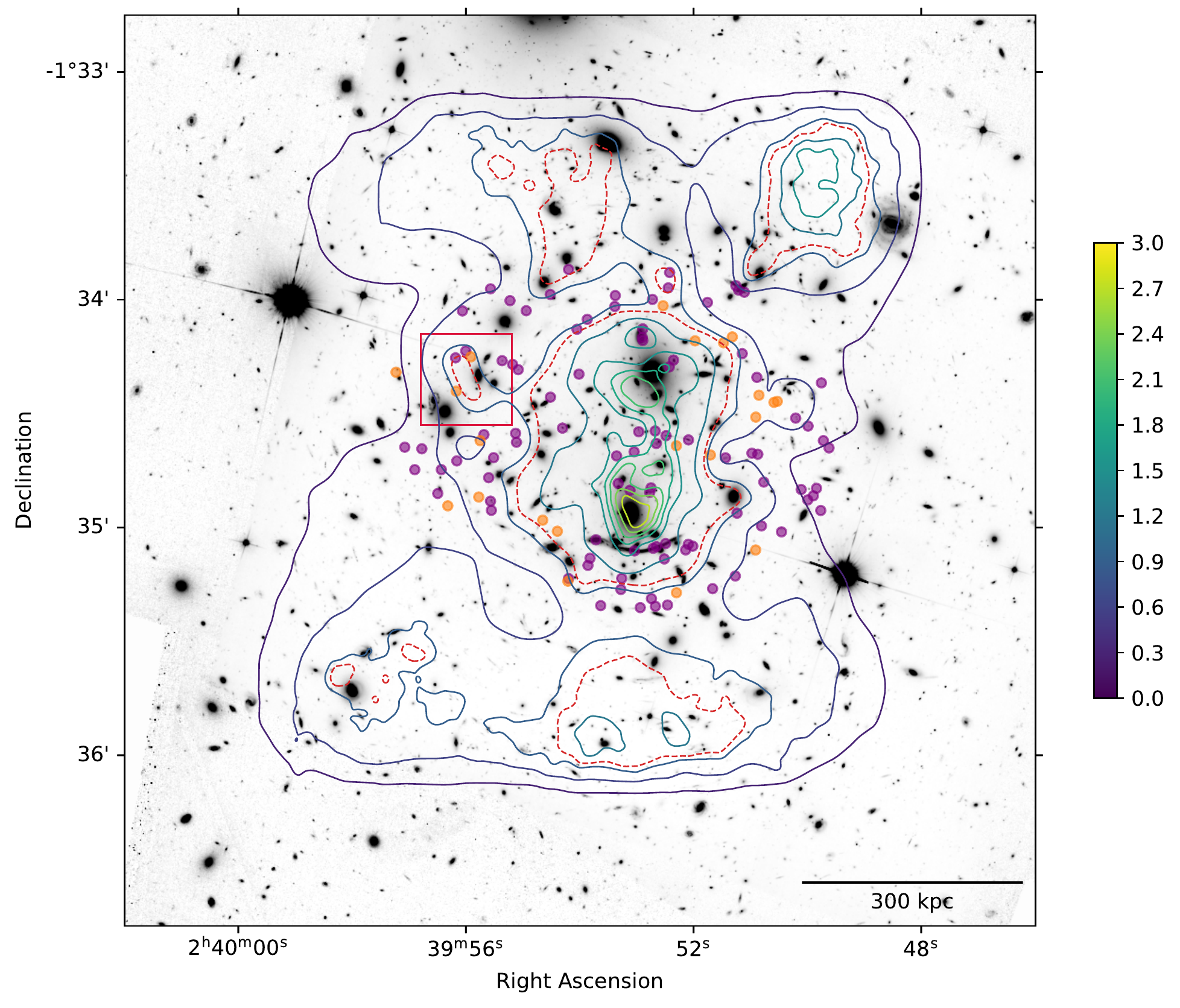}
    \caption{The coloured contours of \grale reconstructed surface mass density distribution of A370 ($z_l=0.375$) overlaid on the BUFFALO mosaic of A370 in the ACS/F814W filter. The surface mass density values are scaled by $\Sigma_{\rm{crit,0}}$. At $z_l=0.375$, 1 arcsec corresponds to approximately $5.15\,$kpc. Red dashed contour correspond to $\kappa=1$. Filled purple dots mark the input images with spectroscopic redshifts (\textsc{gold}) and filled orange dots mark the input images with photometric redshifts (\textsc{silver}).  The region in the red rectangle shows the approximate location of a substructure feature discussed in Section~\ref{sec:substructure}. The mass distributions $\sim 40"$ North and South of the cluster center are discussed in Section~\ref{sec:filament}.}
    \label{fig:massmap}
\end{figure*}

\begin{figure*}
     \centering
     \begin{subfigure}[b]{0.49\textwidth}
         \centering
         \includegraphics[width=\textwidth]{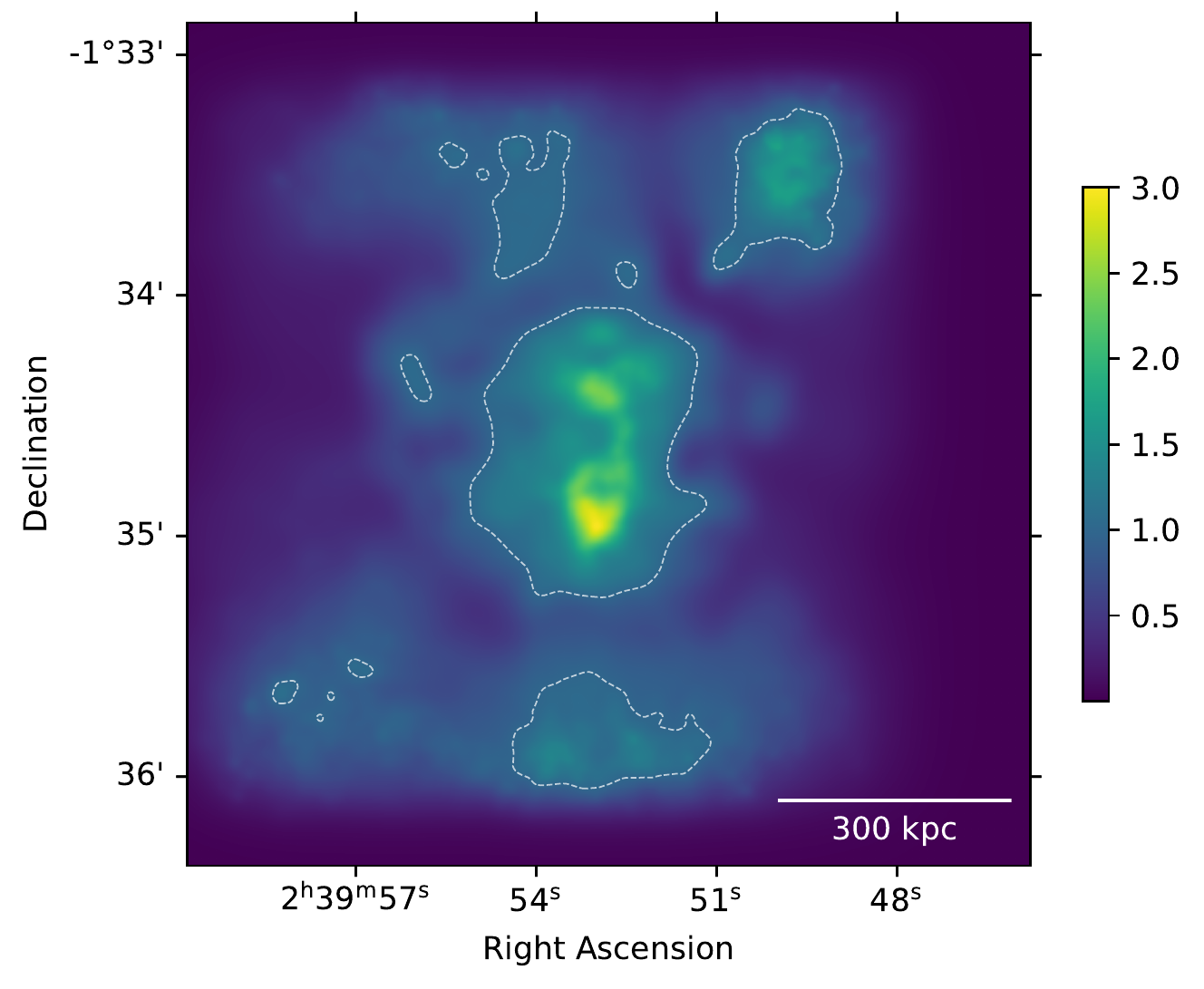}
     \end{subfigure}
     \quad
     \begin{subfigure}[b]{0.49\textwidth}
         \centering
         \includegraphics[width=\textwidth]{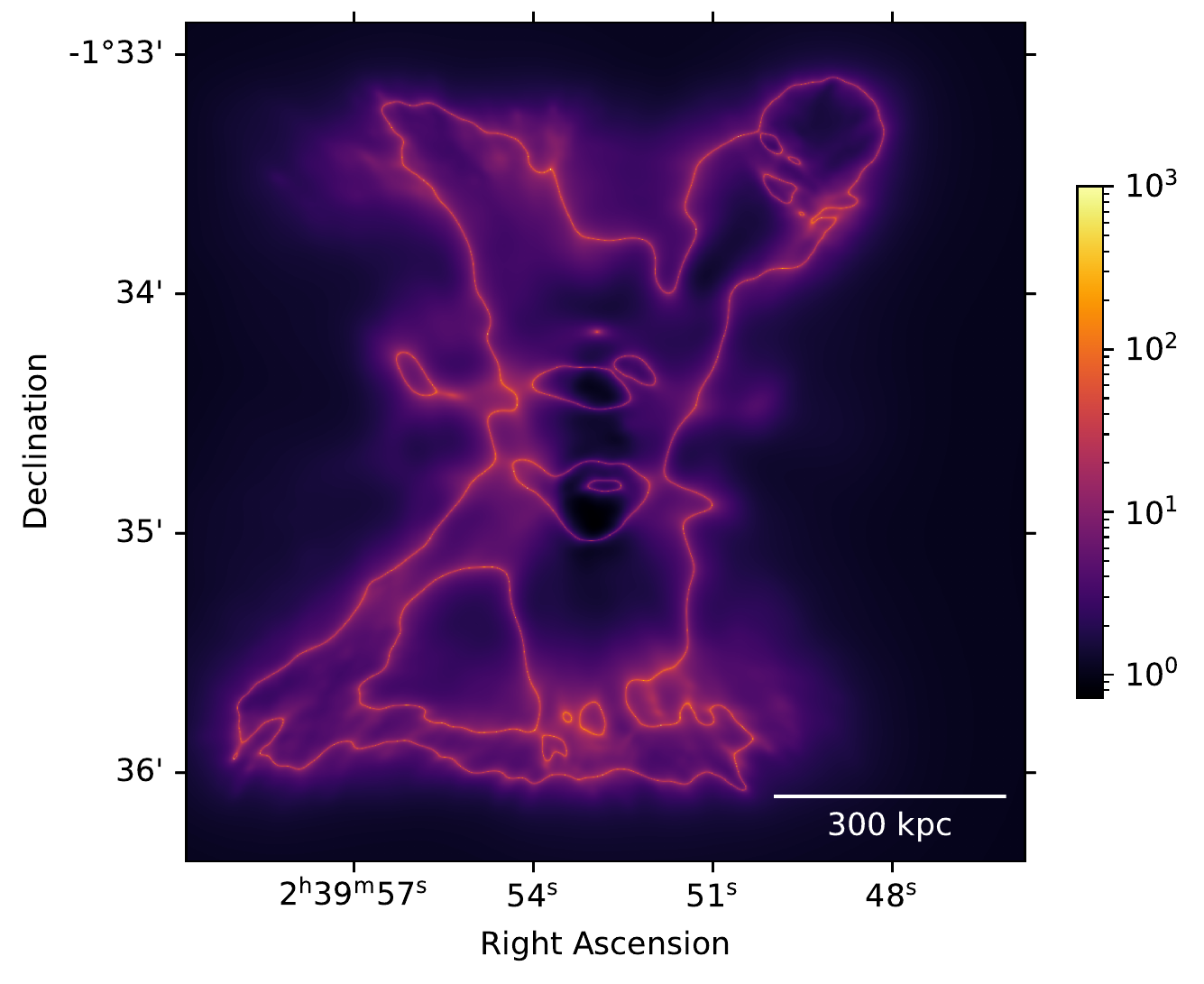}
     \end{subfigure}
     
        \caption{\textit{Left:} \grale reconstructed surface mass density distribution of A370, scaled by $\Sigma_{\rm{crit,0}}$. The dashed white contour represents $\kappa=1$. {\it Right:} \grale reconstructed magnification distribution of A370 in the lens plane ($z_l=0.375$) for sources at $z_s=9.0$.}
        \label{fig:massmapsarcsec}
\end{figure*}

We are using the image data from the \hst~BUFFALO strong lensing multiple image catalog. 
The catalog is made from fully reprocessed \hst mosaics combining the HFF and BUFFALO data, as well as improved spectroscopic data. 
For A370, the catalog consists of 170 strongly lensed images in total.  Spectroscopic data from the Multi-Unit Spectroscopic Explorer Guaranteed Time Observations \citep[MUSE GTO;][]{Lagattuta2017,Lagattuta2019} confirmed spectroscopically 37 systems generating a total of 122 images. The images are visually flagged according to their qualities as {\sc gold}, \textsc{silver}, \textsc{bronze} or \textsc{platinum}, by six different lens modelling groups associated with the BUFFALO collaboration. The \textsc{gold} images are those unanimously flagged as the good images. These are the most secure systems, with optical detection and spectroscopic confirmation. Out of the 37 spectroscopically confirmed systems, 31 are classified as \textsc{gold} systems.  {The other 6 spectroscopically confirmed systems from MUSE do
not have secure identifications on the \textit{HST} mosaics, with some being only tentatively detected or with multiple possible counterparts.}
These are classified as \textsc{platinum} image systems. \textsc{silver} images are good quality images but are a bit less secure than \textsc{gold}, as they do not have spectroscopic confirmation but were still unanimously voted as multiple images. \textsc{bronze} systems are the worst quality images in the catalog. These were not unanimously voted as multiple images and do not have spectroscopic redshifts.  In this paper we are using the  most secured systems, 31 \textsc{gold} systems with spectroscopic redshifts and 8 \textsc{silver} systems with photometric redshifts, which are providing a total of 114 strongly lensed images. The images used in this work are listed in Table~\ref{tab:imgdata}. 

We here list the differences between the image set used for the HFFv4 \grale~reconstructions and the one used in this work updated thanks to the BUFFALO data\footnote{The changes in quality flags are made by a round of voting process by different mass modelling groups within the BUFFALO collaboration. The voting was mainly based on the quality and security of images and their associated spectroscopic information.}:

\begin{itemize}
    \item 4 new \text{silver} systems %have been added to the image catalog 
    - 44, 45, 46 and 56. 
    \item 9 systems with new redshifts assigned to %them - among them the 
    systems 13, 14, 22, 25, 38 and 42 which are identified as \textsc{gold} systems, %with spectroscopic redshifts 
    and systems 8, 11, 41 and 43 as \textsc{silver} systems with photometric redshifts.
    \item 4 systems previously treated as \textsc{gold} and now classified as \textsc{platinum} due to a lack of \hst~optical counterpart but with spectroscopic information from {\it MUSE}. These systems are 32, 33, 34 and 36. These systems are not used in this work.
    \item System 16 has been newly flagged as \textsc{bronze}, and is not used in this work. Image 16.2 is predicted but is not seen in either {\it MUSE} or \hst~data.
    \item Systems 10, 25 and 29 - third images are now flagged as \textsc{bronze} and are not used in this work.
\end{itemize}

The input to \grale consists of the point image locations and redshifts only. In addition to the images, we also used a $250 \times 250$ arcsec null space region (see section 3.1 of G20), to discourage \grale from generating fictitious images in regions of the lens plane where none are observed.

\begin{table*}
\centering
    \caption{Summary of the different inversion methods' mass distribution of the 3 mass features identified in this paper. The works are classified as simply parametrized (SP), hybrid (HY), or free-form (FF), and are arranged approximately chronologically.} 
    \label{tab:3massfeatures}
\begin{minipage}{\textwidth}
\renewcommand{\thempfootnote}{\fnsymbol{mpfootnote}}
    \begin{center}
\begin{tabular}{lcccc}
\hline &                & & Mass feature            &                \\
\hline
Model  &  Type & Mass peak near  &  Substructure    &  N-S filament   \\      
       &       & northern BCG              & 250  kpc east of center       &  (or similar external mass)   \\
\hline 
\citet{Abdelsalam1998}   & FF & Displaced ($\sim 30$  kpc) & Present & Mass extension to the NW \\
\textsc{lenstool} \citep{Richard2010}  & SP &  Displaced ($\sim\!50$ kpc) & Absent & None \\
\textsc{lenstool} HFFv1 \citep{Richard2014}  & SP &  Not Displaced & Absent & None \\
\textsc{Zitrin-ltm HFFv1}   & SP & Not Displaced & Absent  &  None \\
\textsc{Zitrin-nfw HFFv1}   & SP & Displaced ($\sim 35$  kpc) & Absent  &  None \\
\grale HFFv1 \citep{Mohammed2016}           & FF & Displaced ($\sim\!20$ kpc) & Present & Mass clumps near N and S boundaries \\
\textsc{lenstool}  \citep{Johnson2014}  & SP &  Displaced ($\sim\!30$ kpc) & Absent & None \\
\textsc{lenstool} (Sharon HFFv4)  & SP &  Displaced ($\sim\!15$ kpc) \footnote[2]{The nearby halo is elongated NW-SE.} & Absent & None \\
\textsc{lenstool} (CATS HFFv4)         & SP & Not Displaced  & Additional DM halo & Two elongated, low density mass `fingers' \\
\textsc{glafic} \citep{Kawamata2018}   & SP & Displaced \footnote[3]{There are two nearby halos, one is $\sim 20$ kpc away and the other one is $\sim 100$ kpc away.} & Absent & External shear present ($\gamma=6.55\times 10^{-2}$) \\
\textsc{wslap+} \citep{Diego2018b}     & HY & Displaced ($\sim\!40$ kpc) & Present & None \\
\grale HFFv4          & FF & Displaced ($\sim\!15$ kpc) & Present & Mass clumps near N and S boundaries \\
\textsc{swunited} \citep{Strait2018}   & FF & Diffuse ($\sim\!15$ kpc) & Absent & None \\
\textsc{lenstool} \citep{Lagattuta2019}  & SP & Displaced ($\sim\!50$ kpc) & Additional DM halo &  External shear present ($\gamma=1.28\times 10^{-2}$)\\
\textsc{lensmodel} \citep{Raney2020}  & SP & Not Displaced & Absent & None \\
\grale BUFFALOv1 (this work)    & FF& Displaced ($\sim\!35$ kpc) & Present & Mass clumps near N and S boundaries\\
\hline
\end{tabular}
\end{center}
\end{minipage}
\end{table*}

\section{Reconstructed Mass Distribution}
\label{sec:recres}

Our best fit reconstruction is obtained by averaging 40 different and independent \grale runs. Each starts with a random seed. While constrained by the required computational resources, this number is consistent with our previous works \citep[G20;][]{sebesta2019,williams2019}. In this work, we are using a reconstruction area of 0.927 Mpc by 0.927 Mpc and
the smallest resolution cell (projected Plummer sphere) is about 12.5 kpc.
The reconstructed mass distribution, overlaid on a BUFFALO image of A370 using ACS/F814W filter, is shown in Figure~\ref{fig:massmap}. Contours represent the reconstructed projected surface mass density $\Sigma$, scaled by $\Sigma_{\rm{crit,0}}=c^2/ 4 \pi G D_{\rm ol}= 0.314$g/cm$^2$. The same reconstructed surface mass density plot is provided in the left panel of Figure~\ref{fig:massmapsarcsec}. The right panel of Figure~\ref{fig:massmapsarcsec} shows the corresponding magnification distribution in the lens plane. 

It is common to use lens plane rms (LPrms) as an indicator of how well a given mass model reproduces a cluster lens. The LPrms value for this reconstruction is $0.45\,$arcsec. This is an improvement over the earlier reconstruction of the same cluster using HFFv4 data which had LPrms of $0.88\,$arcsec. Since the total number of images used is nearly exactly the same in both cases, the reduced LPrms can be attributed to the improvement of the BUFFALO input data over the earlier HFF data.

Recent reconstructions by most methods in the literature produce low LPrms values, in the range $\sim 0.5"-1"$. Yet the corresponding mass distributions, though sharing many similarities, tend to differ in the details. This is true not just for parametric vs. free-form reconstructions, but among parametric models, even those sharing the same modeling software \citep{Priewe2017}.

\begin{figure}
    \centering
	\includegraphics[width=0.95\columnwidth]{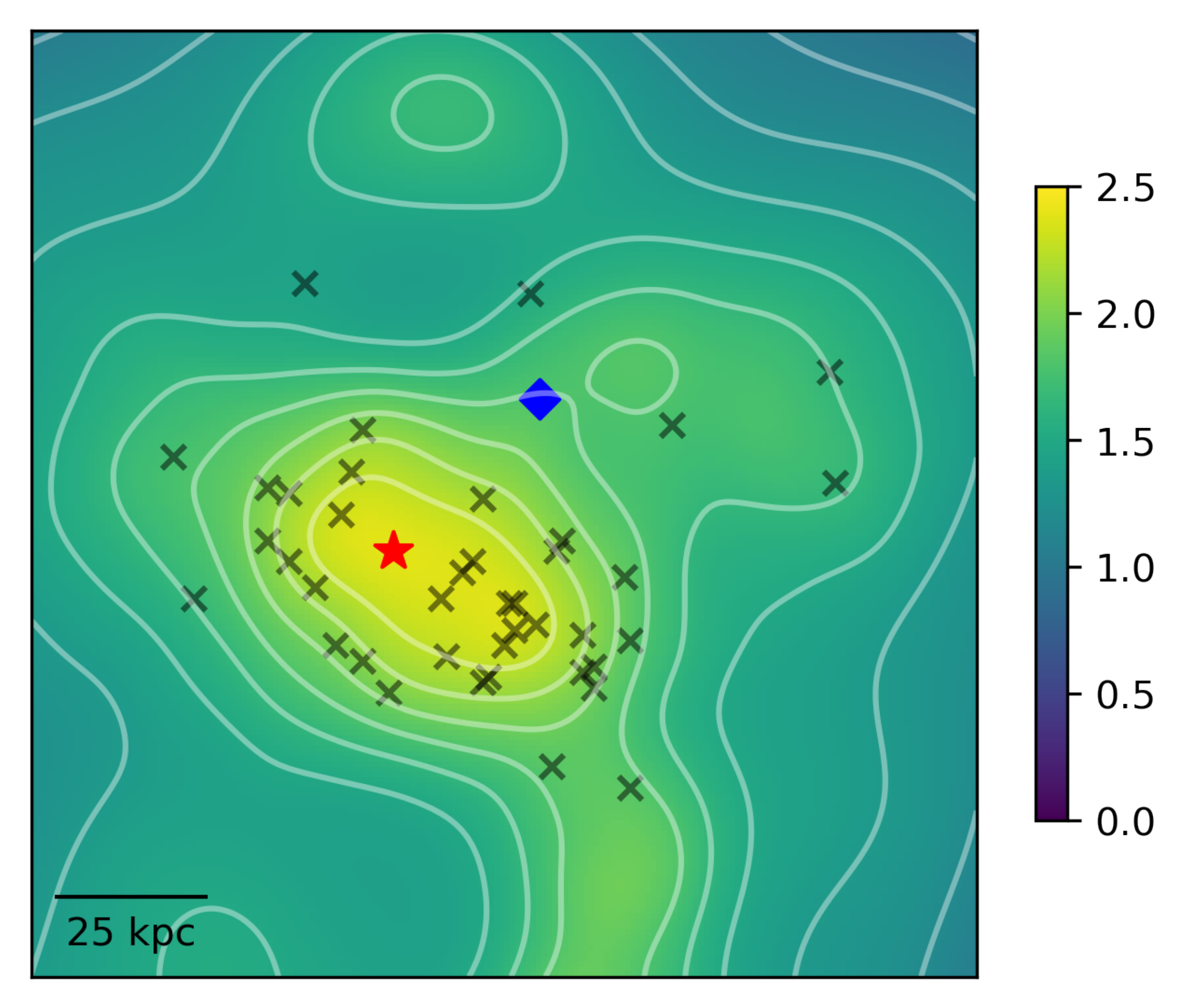}
     \caption{Region of the reconstructed mass distribution highlighting the offset between the northern BCG and the reconstructed northern mass clump. The red star shows the reconstructed northern mass peak from the average \grale map. Position of the northern BCG is marked by the blue diamond. Nearby mass peaks from individual \grale runs are shown as black crosses. The color scale indicates values of $\kappa$ in terms of $\Sigma_{\rm crit, 0}$.}
    \label{fig:bcg}
\end{figure}

\begin{figure*}
	\includegraphics[width=0.8\textwidth]{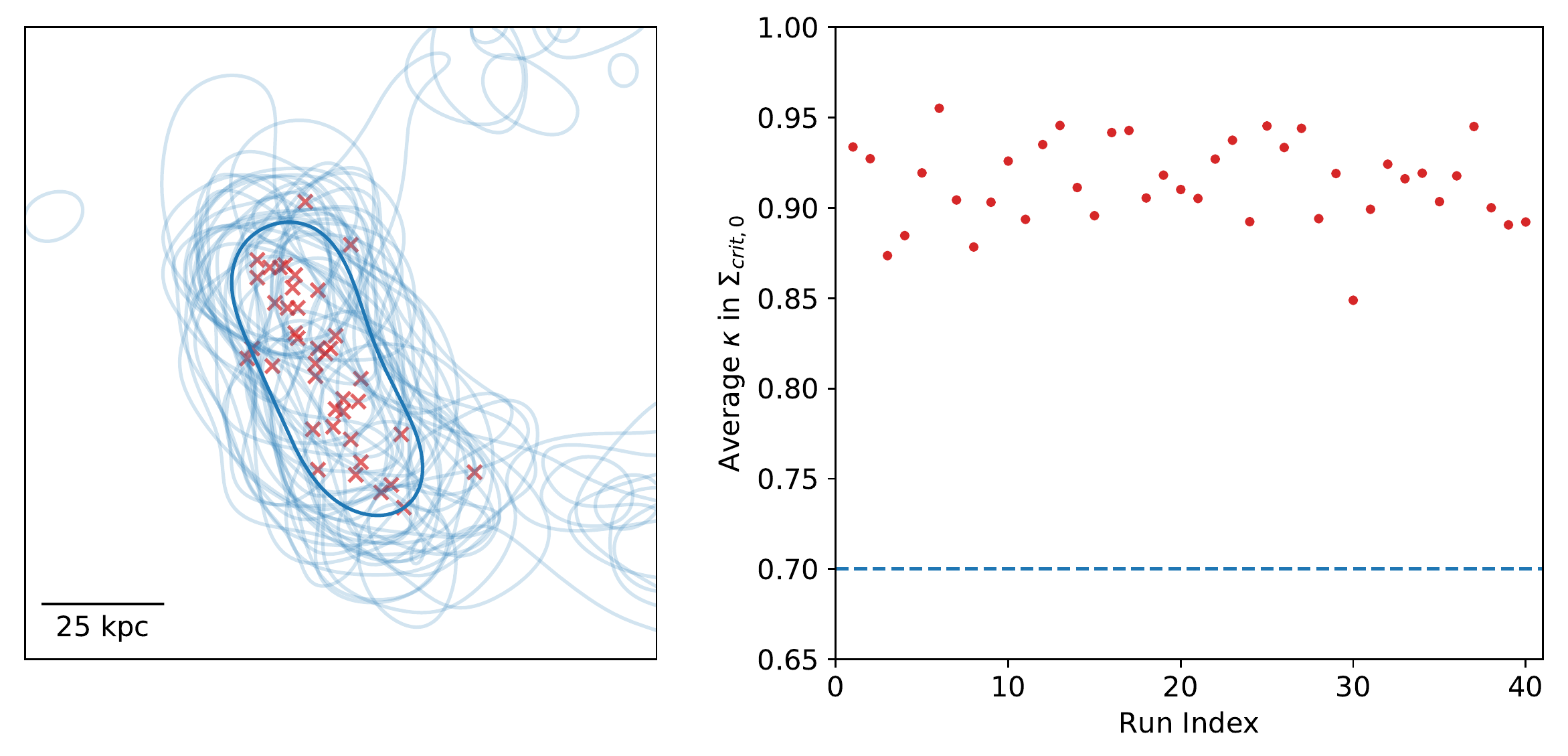}
     \caption{{\it Left:} Region of the reconstructed mass distribution highlighting the $\sim\!10^{12}M_{\sun}$ substructure 250  kpc from the cluster center. Dark blue contour represents the $\kappa=1$ of the average \grale map, where $\kappa$ is in terms of $\Sigma_{\rm crit, 0}$. Light blue contours represent the $\kappa=1$ contour produced by 40 individual \grale runs. Red crosses mark the points with maximum convergence within the substructure, as produced by the individual \grale runs. {\it Right:} Average $\kappa$ values in terms of  $\Sigma_{\rm crit, 0}$ within $\sim100$ kpc region around the substructure, for each individual \grale runs. Dashed blue line shows the average background $\kappa$ around the substructure.}
    \label{fig:substructure}
\end{figure*}

In this section we concentrate on 3 mass features that appear in our resulting mass map: the offset between the northern BCG and the nearest mass peak, the $\sim 100$ kpc mass concentration of super-critical density $\sim 250$  kpc east of the main cluster, and a filament-like structure passing N-S through the cluster. Our reconstructions assume that all the deflecting mass is in the plane of the main lensing cluster. However, it is possible that some of these are foreground or background structures. It is interesting to note that all 3 features were detected as early as \grale\!'s HFFv1 reconstructions \citep{Mohammed2016}, later confirmed by HFFv4 and by the present BUFFALO work. It is even more interesting that all three features also appear to some extent in the first free-form reconstruction of \cite{Abdelsalam1998}, using a different inversion method, although the details of these mass features differ (see their figure 3).

The properties discussed below are summarized in Table~\ref{tab:3massfeatures}.

\subsection{Mass distribution near the BCGs}
\label{sec:bcg}

There are two clear central mass peaks in the reconstruction. The bright central galaxies (BCGs), associated with these central clumps are visible in the background BUFFALO image in Figure~\ref{fig:massmap}. As one can notice, the southern mass clump closely follows the southern BCG on the plane of the sky, but the northern mass clump is significantly displaced from the northern BCG (see Figure~\ref{fig:bcg}). This possible example of mass not following light is consistent with the fact that A370 is an ongoing merger.

The offset we see between the northern BCG and the nearest massive mass clump is broadly consistent with the previous reconstructions performed by a number of groups using the HFF data \citep{Richard2014,Strait2018,Kawamata2018}. In our reconstruction, the displacement is $\sim 35$  kpc South East (SE) of the northern BCG, comparable to that of the \cite{Lagattuta2019} model, where the center of their DM3 halo is displaced by $\sim 50$ kpc somewhat East of SE from that BCG.  However, in another parametric model, \cite{Kawamata2018}, there are two cluster-scale halos near the northern BCG, one $\sim 20$ kpc to the south, and the other located $\sim 100$ kpc SE away from the northern BCG.  Even though all models agree that there is significant mass not directly associated with the BCG to the SE of it, not all agree on the location, to within $\sim 50$ kpc. 

Since all models reproduce the image distribution very well, it is the lensing degeneracies (as well as the differences in input data) that lead to these disagreements. In this particular case, the main degeneracy is most likely the monopole \citep{Saha2000,Liesenborgs2012}, because the $\sim 120$ kpc region just SE of the northern BCG is devoid of lensed images, allowing any number of monopole-like redistribution of mass.

\subsection{A $\sim\!10^{12}M_{\sun}$ substructure 250  kpc from cluster center}
\label{sec:substructure}

A recent model comparison paper \citep{Raney2020a} analysed half-sample mode of the magnification maps of the 6 HFF clusters. Half-sample mode is the maximum likelihood value of the magnification distribution at every pixel in the lens plane. They identified a mass feature in A370 that was independently recovered by a free-form \grale\!, and a hybrid \textsc{wslap+} method (see their figure 17 and Section 4.1.1). It is enclosed by the red rectangle in Figure~\ref{fig:massmap}. 

This feature is absent in most HFF parametric reconstructions, including the more recent ones \citep[e.g., Sharon HFF v4,][]{Kawamata2018,Raney2020a}, as there is no light associated with it. Also, in \cite{Strait2018}, which used the free-form code \textsc{swunited}, the substructure is not seen. However, CATS HFF v4, \cite{Lagattuta2017} and \cite{Lagattuta2019} include it as an additional cluster-scale dark matter component, at a location very similar to where \grale places it. This disagreement between recent reconstructions cannot be entirely due to the lack of lensing constraints: unlike the regions SE of the northern BCG, the $r\sim 50$  kpc ($\sim 10$ arcsec) region centered on this mass clump has roughly 5 images. These variations in the mass reconstructions in this region are an example of lensing degeneracies, and not just between free-form or hybrid and parametric models, but among parametric models -- CATS HFF v4 and Sharon HFF v4 -- that use the same \lenstool software.

Our BUFFALO reconstruction also recovers this feature, giving more credence to its reality. It is the only feature with $\Sigma>\Sigma_{\rm crit,0}$ outside of the central cluster, and mass clumps described in Section~\ref{sec:filament}. The average surface mass density within that $\sim100$ kpc region is $0.9 \Sigma_{\rm{crit,0}} (\approx 0.3$ g/cm$^2)$, a factor of 1.3 larger than that in the surrounding regions. The total excess mass is $2.6 \times 10^{12} M_{\sun}$. Its relatively large mass is probably the reason why it was already detected more than 20 years ago, by a different free-form lens inversion method \citep{Abdelsalam1998}.  All the individual \grale runs have produced this substructure. The distributions of mass within and around the substructure, as produced by individual \grale runs, are presented in Figure~\ref{fig:substructure}.

\subsection{A North-South Filament-like Structure}
\label{sec:filament}

Our reconstructed mass distribution (Figure~\ref{fig:massmap}) contains significant mass clumps, located $\gtrsim 40$ arcsec ($\gtrsim 200$ kpc) above and below the central region, which obviously do not represent actual mass at those locations. These fictitious mass clumps are generated by \grale in order to minimize the fitness values for the given set of input images, i.e., to reproduce the observed image positions as best as possible.
This can be attributed to \grale trying to compensate for an otherwise unobserved mass outside the strong lensing region, which, as we illustrate below, is consistent with a filament-like structure passing through this cluster. These probably lie well outside the strong lensing region, and must be sufficiently massive to influence the deflection angles of the lens in the central part, and thus the strongly lensed image locations. 

Though no other reconstruction of A370 has such fictitious masses, many include features that result in similar influence on the image locations. CATS HFF v4 reconstructed mass map  includes two highly elongated, low density mass `fingers' with an angle of $\sim 40$ degrees between them, arranged in an X-like pattern, centered roughly on the cluster center and extending North-South well beyond the strong lensing region (see their mass models on MAST\footnote{\url{https://archive.stsci.edu/prepds/frontier/abell370\_models\_display.html}}). We suspect that these have the same role as our external mass clumps. 

In the \textsc{glafic} reconstruction of \citet{Kawamata2018}, the authors include an external shear of $\gamma=6.55^{+2.25}_{-2.35}\times 10^{-2}$ with a position angle of $\theta_{\gamma}=177.71^{+4.13}_{-5.64}$ degrees at a redshift $z=2$ (see their table 8). External shear is often used to account for mass located outside the modeled region. Its orientation is consistent with our speculation of the presence of outlying mass structures, above and/or below the central region of the cluster. 

\cite{Lagattuta2019} model also has external shear, $\gamma=0.128$, whose inclusion considerably decreases the model's LPrms from $\sim 1$ to $0.66$ arcsec. {To replace the external shear with a more physically motivated component}, the authors carry out a useful exercise of figuring out what mass distribution could be responsible for the shear. They test a few possibilities and conclude that the mass associated with the observed background, foreground, and in-cluster galaxy groups are not responsible for the shear. This leaves mass further afield as the only possibility, consistent with our conclusions.

It is interesting to recap how \grale and the \lenstool\!\!-based parametric analysis of \cite{Lagattuta2019} localized the mass responsible for the external shear. \grale\!, being sensitive to images only, must have picked up on the small systematic image deflections, tracing their source to outside (above and/or below) the cluster's central $\sim 400$  kpc region. The parametric model, on the other hand, had to try a number of possible sources of mass and rule them out one by one if they did not produce a better fit. For a discussion on how deflection fields are affected by external mass clumps, see the work of \cite{Mahler2018} on the cluster Abell 2744.

\citet{Lagattuta2019} also pointed out the possibility of additional mass structures in the outskirts of A370 as a replacement to their external shear. In their figure 11 they showed several concentrations of galaxies with similar colors to cluster members, in lower resolution image taken with the Canada-France-Hawaii Telescope (CFHT) with CHF12K \citep{Hoekstra2007}. The blue contours in that figure shows smoothed out light distribution from the cluster red sequence members. One can see mass structures towards North North West and South of SE, and within the virial radius. 
Spectroscopic redshift measurements for a few of these objects by early CFHT/PUMA and ESO/PUMA2 \citep{Fort1986} placed them at the same distance as the cluster \citep[see figure 1 in][]{Mellier1988}.  The size and locations of these structures are promising as a source for the additional structures \grale has produced.

\begin{figure*}
	\includegraphics[width=0.8\textwidth]{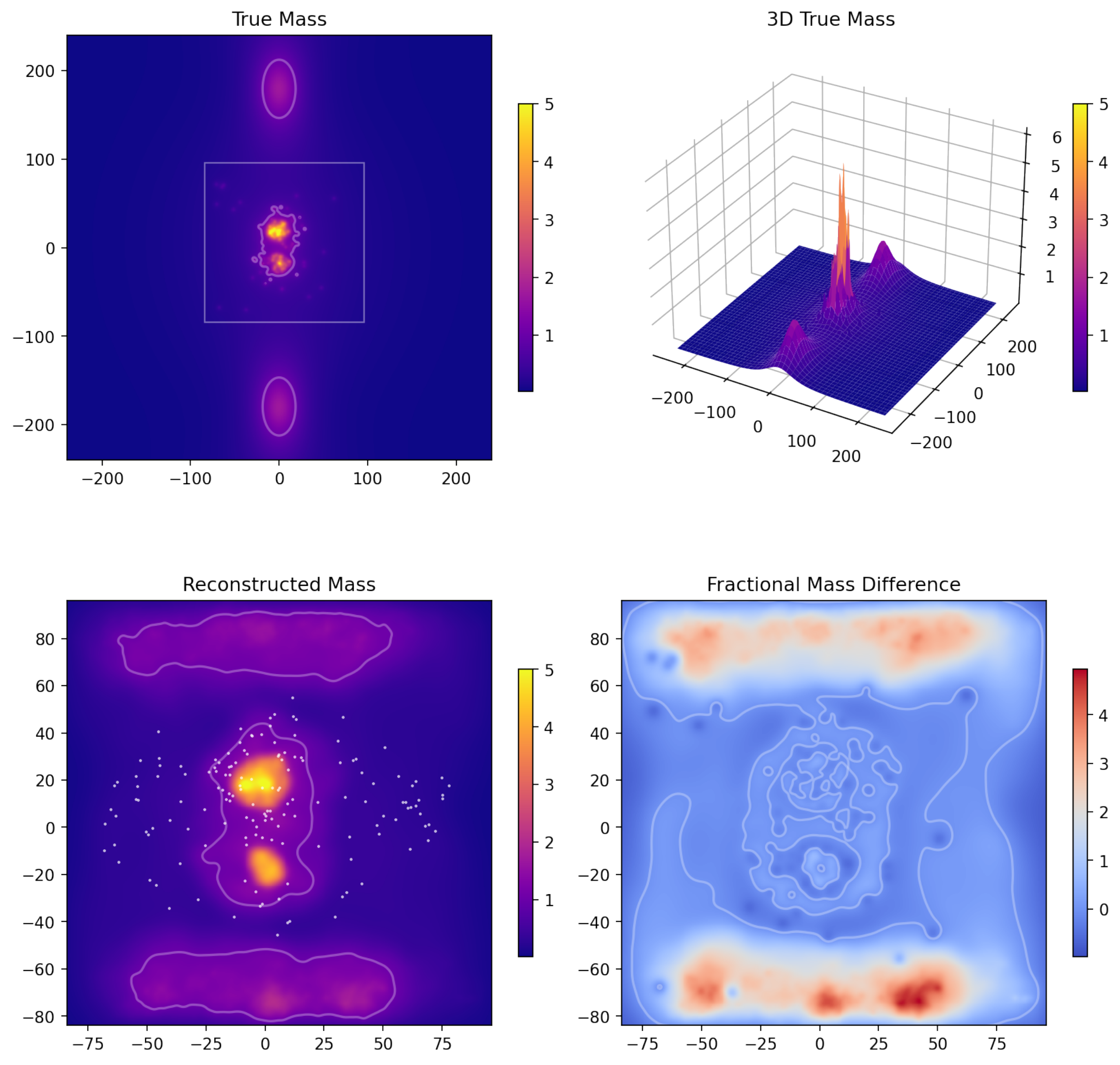}
    \caption{Mass distribution maps for the synthetic cluster Irtysh IIIc. \textit{Upper left:} True surface mass density distribution. The white box represents the region used for reconstruction. There are two outer mass clumps lying well outside this box. Faded white lines represent the $\kappa=1$ contour.  \textit{Upper right:} Three dimensional presentation of the true surface mass density distribution. \textit{Lower left:} Reconstructed surface mass density distribution. White points represent the input images used in \grale, and the faded white lines represent the $\kappa=1$ contour. \textit{Lower Right:} The fractional mass difference (${\Delta m}/{m}$, as defined in Equation~\ref{eq:fdiff}) map between the true and the reconstructed surface mass densities. Faded white contours represent the ${\Delta m}/{m}=0$ contour. According to the color scheme overestimated mass is represented as more red and underestimated mass as more blue.
    In all of these maps the surface mass density values are scaled by $\Sigma_{\rm{crit,0}}$.} 
    \label{fig:irtyshiii} 
\end{figure*}

In the following subsection we present the \grale analysis using a synthetic cluster, which reinforces the conclusion that A370 likely has a filament-like structure passing North-South through/near it.

Our conclusion from considering the 3 mass features described above is that many models agree on these (though not necessarily down to fine details) in the reconstruction of A370. The interesting point is that the models that agree with each other are not all of the same modeling type (for example, not all are parametric), but span the range of modeling philosophies: parametric, hybrid and free-form. At the same time, some models of the same type, and even the same modeling software, disagree with each other.

\subsection{Modelling a synthetic cluster with filament: Irtysh III}
\label{sec:irtyshiii}

Our hypothesis for the origin of the fictitious mass clumps towards the top and bottom edges of the reconstructions (Section~\ref{sec:filament}) is that these replicate the effect of actual mass structures lying well outside the reconstructed region. The presence of these masses minimizes the fitness value of the overall reconstruction, i.e., better reproduces the observed images. To verify this hypothesis, we performed a reconstruction of a synthetic cluster which mocks a situation with filament-like structures placed well outside of the cluster strong lensing region.

\subsubsection{Method}
\label{sec:irtyshmethod}

In G20, we worked with the synthetic cluster Irtysh to show that increasing the number of input images improves the quality of the lens reconstruction, and has the potential to predict the Hubble's constant with $\lesssim 1$ per cent precision. On the contrary, the lens plane rms value increases, counter-intuitively, due to the complexity of fitting the increasing number of constraints from strongly lensed images.

Since we had two other versions of Irtysh, Irtysh I and II, thoroughly examined in G20, the version of Irtysh used in this paper is named Irtysh III.

For simplicity and computational efficiency, the mass distributions for these mock galaxy clusters were generated using the analytic softened power-law ellipsoid potential called {\tt `alphapot'} from the \textsc{gravlens} catalog of models \citep{Keeton2001}, to represent the cluster-scale and the galaxy-scale projected lensing potential,
\begin{equation}
    \Psi=b(s^2+\xi^2)^{\frac{\alpha}{2}},
    \label{eq:alphapot}
\end{equation}
where $b$ is the normalization, $s$ is the core radius that eliminates the central singularity, and $\xi^2=x^2+y^2/q^2+K^2xy$ with $q$ and $K$ together representing ellipticity with non-zero position angle. Because the cluster is synthetic, this mass distribution can be rescaled to any size. 

The advantage of using an analytic potential is that the values of the deflection angles and the surface mass density can be determined exactly. We also note that the profiles we use to build up our synthetic clusters i.e. the elliptic potential {\tt alphapot}, is different from the basis functions {\grale} uses for reconstruction, which are projected Plummer spheres. 

Irtysh III is made as a superposition of two massive cluster-scale dark matter components, and $115$ galaxy-scale components. The number and normalizations of the latter are determined by assuming the mass function for clusters of galaxies \citep{Schechter1974,Bahcall1993}. In order to mock a cluster with outlying masses, two identical large elliptical mass clumps, each $88.1\%$ of the central cluster mass, were placed at a distance of $180$ arcsec ($0.967$ Mpc) from the cluster center, above and below the central region. They remain significantly outside the strong lensing region that \grale uses in reconstructions. The true projected mass distribution of Irtysh III is shown in the upper panel (left: two dimensional; right: raised relief map) of Figure~\ref{fig:irtyshiii}. 

\subsubsection{Input}
\label{sec:irtyshinput}

We assumed the cluster to be at a redshift of $z_{\rm l}=0.4$. We are using 151 input images for the reconstruction of Irtysh III. This means this is the `c' type of reconstructions following the notation used in G20. The `a' and `b' types of reconstructions are those with about $1000$ and $500$ input images, respectively. The images were produced following the procedure described in Section 2.3 of G20.

\subsubsection{Results}
\label{sec:irtyshresults}

The reconstructed best fit mass distribution is obtained by averaging 40 different and independent \grale runs, shown in the lower left panel of Figure~\ref{fig:irtyshiii}. The colour-bar represents the projected surface mass density $\Sigma$, scaled by $\Sigma_{\rm{crit,0}}=c^2/ 4 \pi G D_{\rm l}= 0.314$g/cm$^2$. As per our expectations, \grale has produced fictitious mass clumps above and below the cluster center, in regions with no observed images. \grale created these clumps to replicate the effect of true mass structures lying outside the reconstructed region, in order to minimize the fitness of the reconstruction. The lens plane rms of this reconstruction is  0.21 arcsec.

If the fictitious outer mass clumps in the reconstructions are mimicking the deflection field produced by the filament-like mass clumps in the true Irtysh III, we expect the masses and distances of the two sets of clumps to scale according to the scaling relation of external shear.
 
In the true mass distribution, each of the upper and lower outer masses are about $88.1\% $ of the central cluster mass. They are at a distance of $180$ arcsec from the cluster center,  whereas in the reconstruction the upper and lower mass clumps are at a distance of $\sim 80$ arcsec. The external mass contribution to the deflection field in the central region varies as $\sim M/r$, where $r$ is the distance of the external mass clumps to the cluster center, and $M$ is the external mass. Therefore, one would expect the amount of mass in the upper or lower clumps in the reconstruction to be approximately, $88.1\%\times \frac{80 {\text{\rm arcsec}}}{180 {\text{\rm arcsec}}} \simeq 39\%$. The average reconstructed masses in the upper and lower clumps are about $63\%$ and $57\%$ of the reconstructed central cluster mass, respectively. This is fairly comparable to our estimate above, indicating that \grale is sensitive to mass outside of the image region, and replaces distant mass clumps with appropriately scaled masses inside the reconstruction box, but outside the image region.

In the lower right panel of Figure~\ref{fig:irtyshiii}, the fractional mass difference map is shown. This quantity is defined as, 
\begin{equation}
\frac{\Delta m}{m}=\frac{m_{\rm reconstructed}-m_{\rm true}}{m_{\rm true}},
\label{eq:fdiff}
\end{equation}
The value of this quantity determines whether \grale has overestimated or underestimated the mass in a certain region. Clearly, near the fictitious mass clumps the fractional mass difference is quite high. On the contrary, it is small near the central and smaller mass peaks of the true mass distribution. For the very small true mass peaks, \grale does not have enough constraints and/or enough angular resolution to resolve them separately.

The fictitious mass clumps in the reconstruction of Irtysh III look very similar to those in A370, which strengthens our hypothesis that A370 has a filament-like structure running North-South through it.

\section{Reconstructed Magnifications}
\label{sec:mag}
\subsection{Magnification distribution in the source plane}
\label{sec:magsp}

\begin{figure*}
	\includegraphics[width=0.75\textwidth]{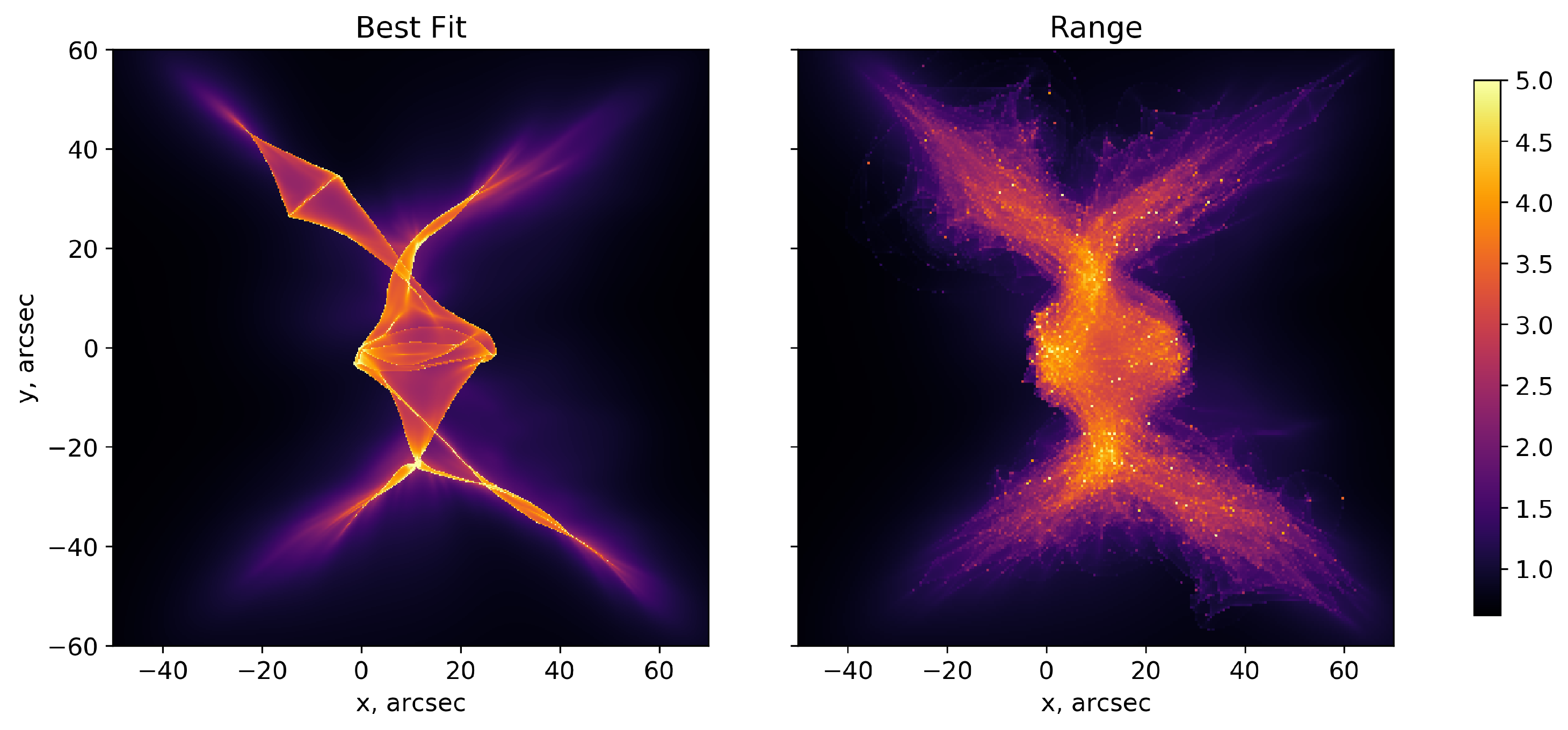}
    \caption{Magnification maps for A 370 in the source plane at a redshift of $z_{\rm s}=9.0$. The left panel shows the best fit magnification maps i.e. magnification values are generated from the average of all the 40 \grale runs. The right panel shows the normalized superposition of the range distribution of magnifications in individual \grale runs in the source plane. In other words, the left panel shows the magnification of the average and the right panel shows the average of magnifications.} 
    \label{fig:A370magmaps} 
\end{figure*}

\begin{figure*}
	\includegraphics[width=0.75\textwidth]{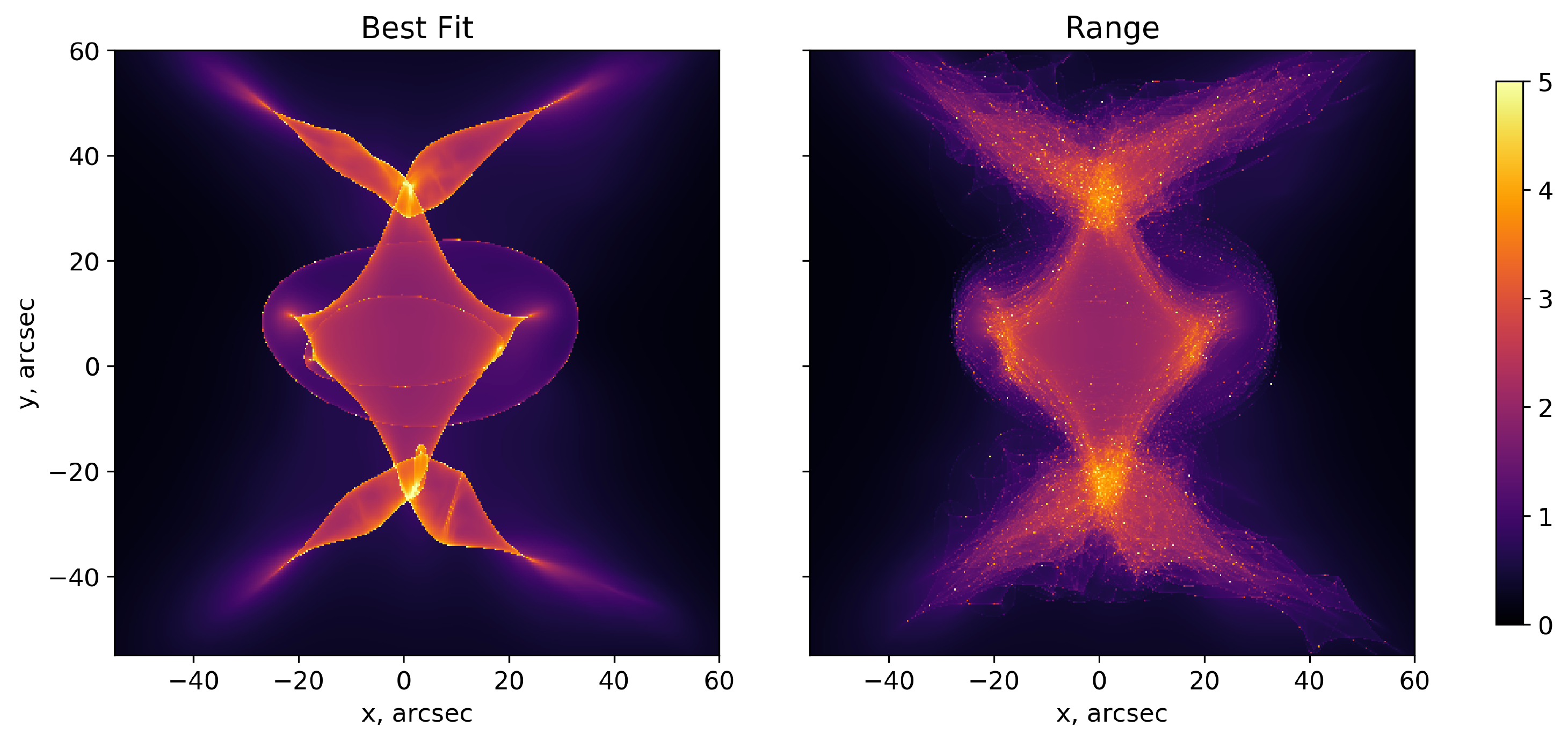}
    \caption{Same as Figure~\ref{fig:A370magmaps} but for synthetic cluster Irtysh IIIc; see Figure~\ref{fig:irtyshiii}.}
    \label{fig:irtyshiiimag} 
\end{figure*}

\begin{figure*}
	\includegraphics[width=0.75\textwidth]{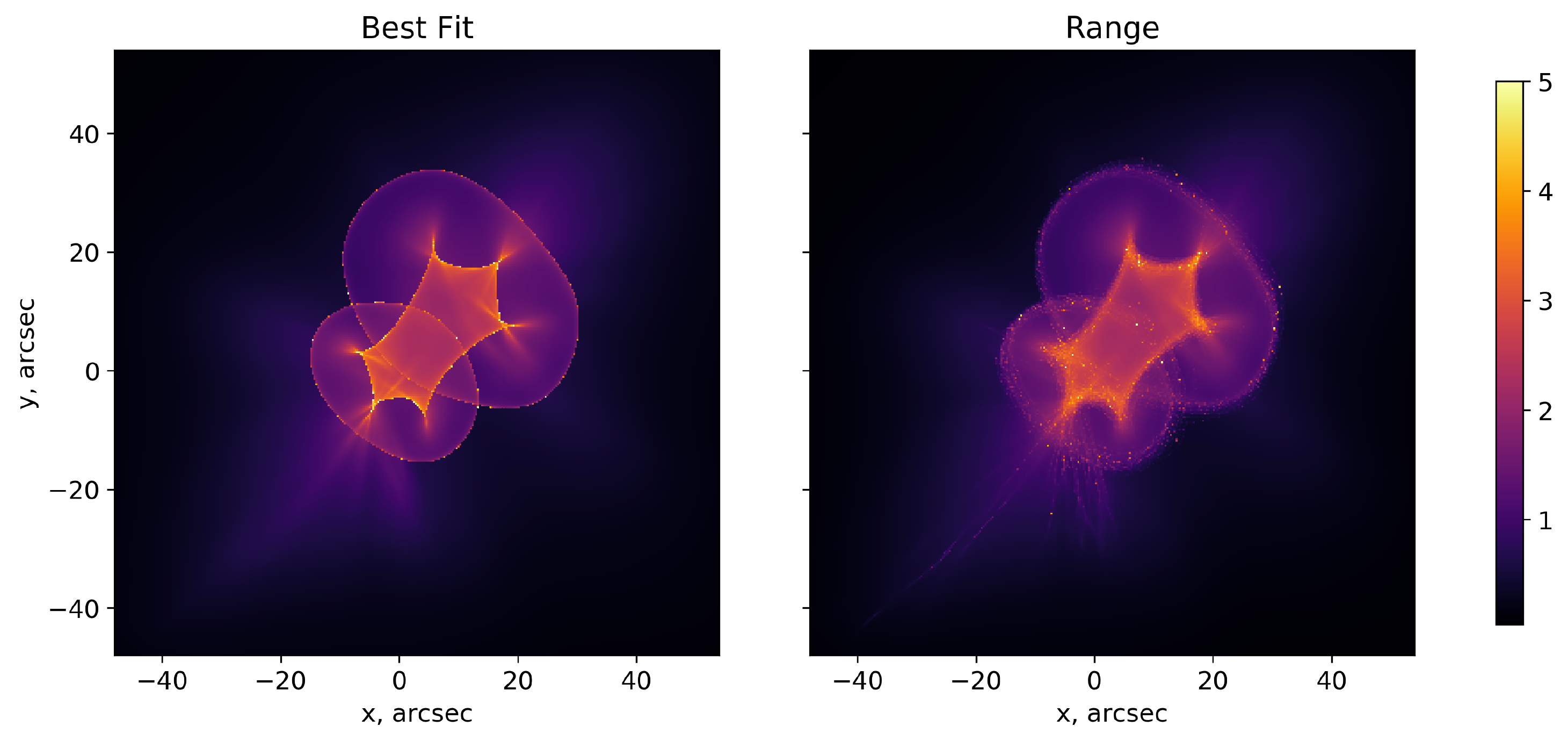}
    \caption{Same as Figure~\ref{fig:A370magmaps} but for synthetic cluster Irtysh Ic (see G20).} 
    \label{fig:irtyshimagmaps} 
\end{figure*}

\begin{figure*}
     \centering
     \begin{subfigure}[b]{0.47\textwidth}
         \centering
         \includegraphics[width=\textwidth]{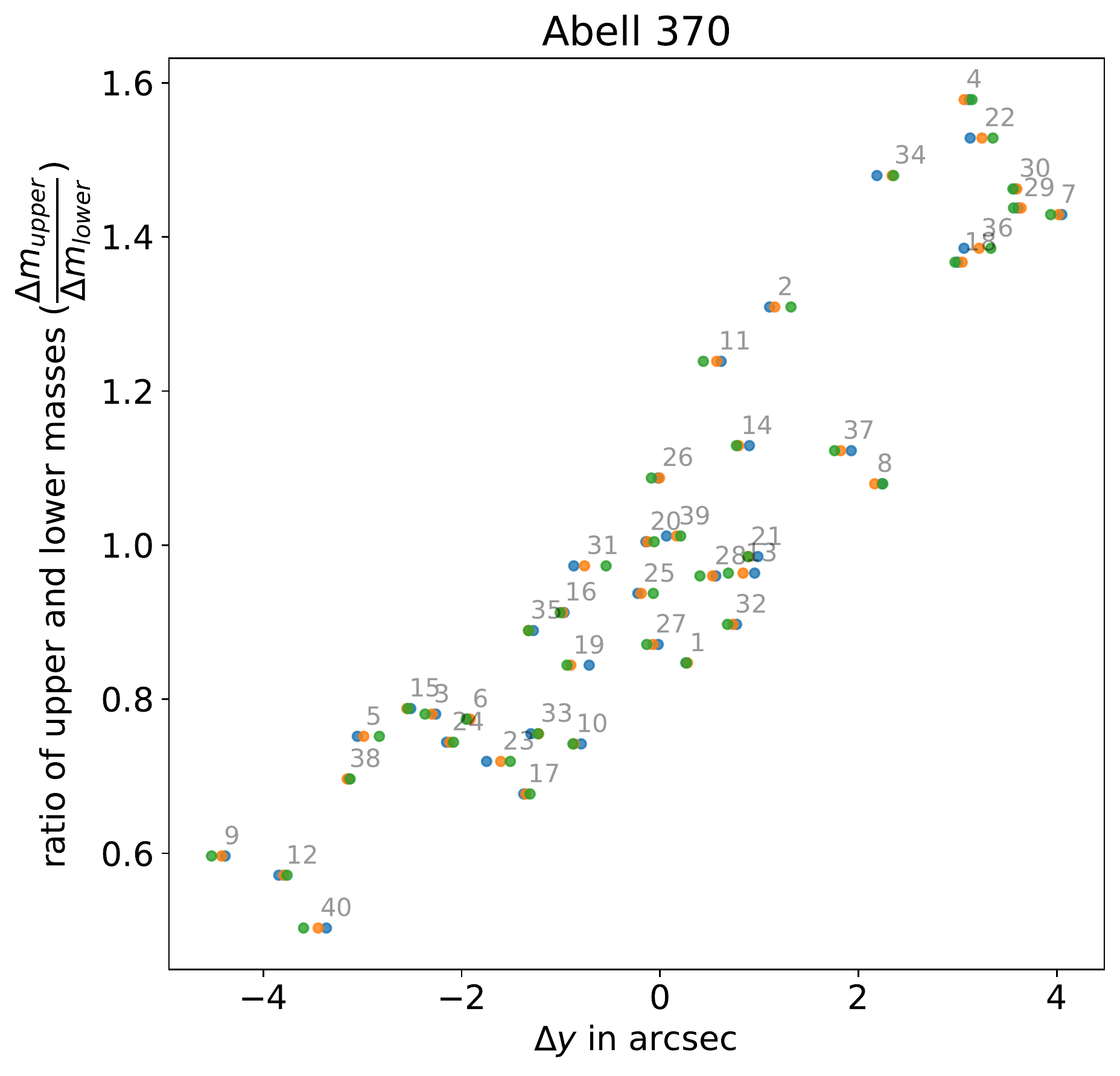}
     \end{subfigure}
     \quad
     \begin{subfigure}[b]{0.47\textwidth}
         \centering
         \includegraphics[width=\textwidth]{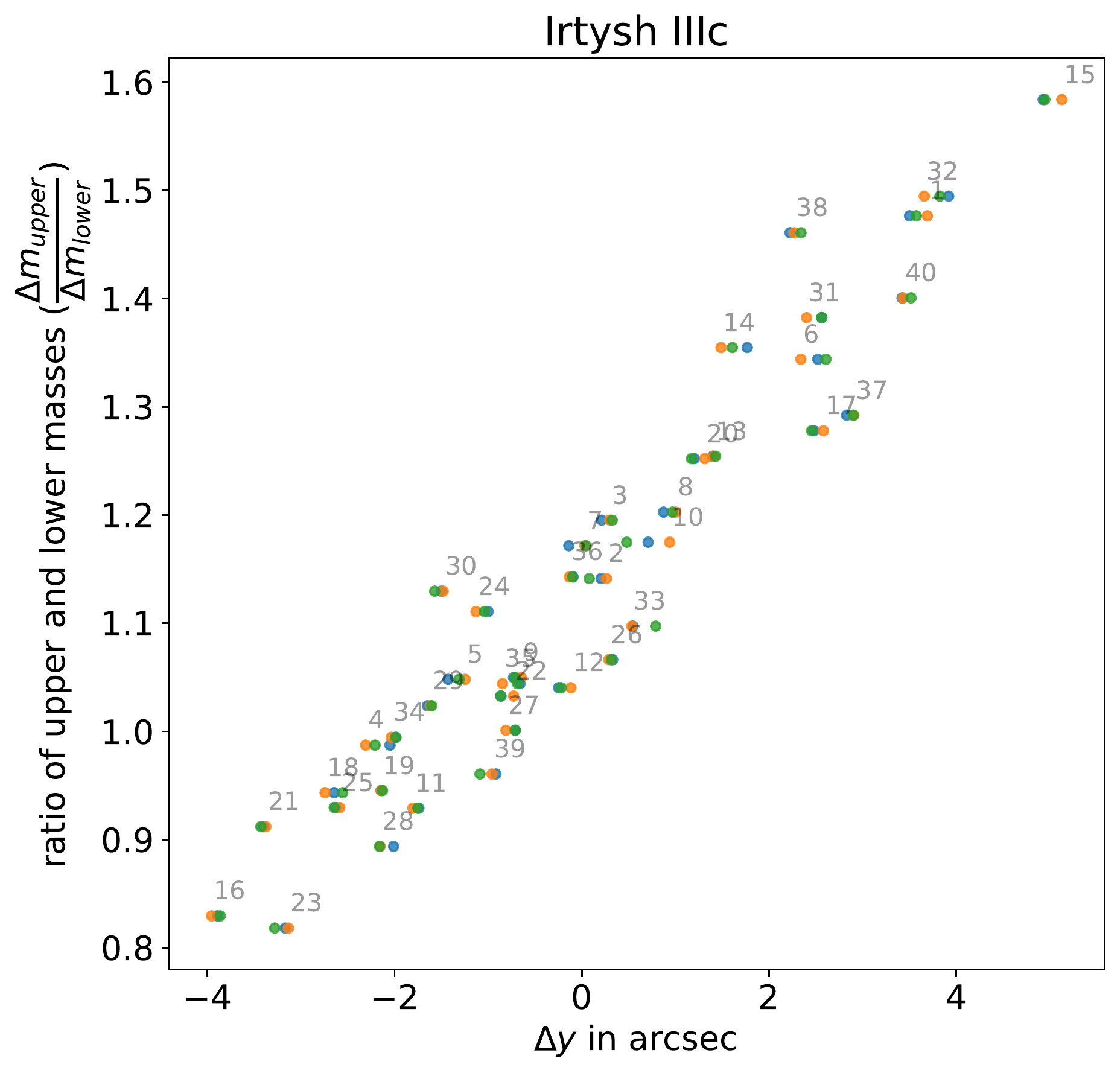}
     \end{subfigure}
        \caption{Ratio of the outer upper and lower masses as a function of the vertical shifts of the back projected images for individual \grale runs for A370 (left) and Irtysh IIIc (right) reconstructions. We are considering only the first source in the catalogs for each of the clusters. In both cases there are three multiple images each represented by different colors.  The numbers annotated beside the the set of points are representing the 40 \grale runs. }
        \label{fig:vertshift}
\end{figure*}

To calculate the magnification distribution in the source plane of A370, we used the best fit reconstruction, which is an averaged combination of all the 40 individual \grale runs. We generated an $120\times 120$ arcsec grid in a source plane of a given redshift, with a grid spacing of 0.25 arcsec. Considering each of the grid points as a point source, we forward lensed them using the reconstructed deflection angles and the lens equation. In this way, we generated one or multiple images for each of the points in the source plane grid. Then we calculated the sum of the unsigned magnifications for all the images in the image plane, generated from a single source point, and assigned it to be the magnification of that corresponding source. This produces the reconstructed magnification distribution in the source plane.

We want the readers to note that our method of calculating the magnification distribution in the source plane differs from the source plane and the image plane methods described in \citet{Diego2019} and \citet{Vega2019}. 
For a comparative discussion, see Appendix~\ref{sec:magcomp}.

The left panel in Figure~\ref{fig:A370magmaps} shows the best fit magnification distribution in a source plane at a redshift of $z_{\rm s}=9.0$. One can easily notice the caustics i.e. the contours of high - theoretically infinite - magnifications.

Since we are using an average of 40 different \grale runs, magnification maps can also be obtained for each of these \grale runs individually, giving rise to the reconstructed range of the source plane magnification distributions. The right panel of Figure~\ref{fig:A370magmaps} plots a normalized superposition of the 40 individual A370 reconstructions.
In this range of maps, one can see the caustic features from all the individual runs. 

An interesting outcome of this exercise is shown in the right panel of Figure~\ref{fig:A370magmaps}: in the normalized superposition of range magnifications maps the caustics from 40 individual runs are not completely overlapping with each other. Rather, they are shifted significantly either towards the top or the bottom of the source plane for different runs, creating an overall fuzzy appearance. The main reason for these shifts is that, for different individual runs, \grale produces different amounts of extra mass in the clumps towards the northern and southern edges of the central region. 

Earlier we noted the presence of these mass clumps in the best fit maps of A370, and argued that they are due a filament-like structure outside of the modeling window.  We now see that individual maps distribute the mass differently between northern and southern edges. It is interesting to compare this with the synthetic cluster Irtysh III, which was constructed with filament-like structures north and south of the lens center.

In the left panel of Figure~\ref{fig:irtyshiiimag}, we show the best fit magnification distributions in the source plane of Irtysh III. As in the case of A370, the reconstructed range distribution of the source magnification maps shows that the caustics are shifted up and down for individual \grale runs (right panel of Figure~\ref{fig:irtyshiiimag}). This similarity between the reconstructions of A370 and Irtysh III strengthens our hypothesis that the extra outlying mass clumps are due to filament-like structures above and/or below the main cluster. It is also worthwhile to mention that for the \grale reconstructions of synthetic Irtysh Ic - where no external filament-like masses were present and no extra mass clumps were created by \grale (see G20) - all the caustics in the range magnification maps are superimposed, unlike those in Irtysh IIIc and A370 (see Figure~\ref{fig:irtyshimagmaps}).

\subsection{Relation between outer mass clumps and back-projected images}
\label{sec:vert}

In the previous subsection, we concluded that individual \grale runs produce different amounts of extra mass towards the northern and southern edges of the central region, causing vertical shifts in the caustics. For a further verification of this argument, we calculated the vertical shifts, $\Delta y$, for each of the back-projected images in the source plane, and for each individual run, from their averaged back-projected positions in the best fit reconstruction. Caustics trace out the source plane positions with theoretically infinite magnifications. Therefore, the vertical shifts of back-projected images (which are basically the reconstructed source positions) are directly related to the shifts of the caustics. If our hypothesis holds, $\Delta y$ should be related to the ratio of the amounts of mass lying outside $\pm 50$ arcsec from the cluster center, in the northern and southern directions.

In the left panel of Figure~\ref{fig:vertshift}, we plot the vertical shifts of the back projected images against this mass ratios for each of the 40 \grale runs of our A370 reconstructions. The plot shows only the first source with three images. Each of these three back projected images are marked by three different colors, red, green and blue. The annotated numbers by the sides of the plotted points mark the corresponding \grale run number. As one can see, $\Delta y$ is more positive when the northern edge has a greater amount of extra mass (the mass ratio is higher), and is more negative when the southern edge has more mass. The rest of the sources exhibit similar trends. These clearly imply that \grale generates degenerate solutions for individual runs by varying the fraction of mass in the upper vs. lower mass clumps.

Now, it is interesting to check whether the vertical shifts of the caustics and ratio of the upper to lower masses are also related in the same fashion for Irtysh III. This is shown in the right panel of Figure~\ref{fig:vertshift}, for the case of the first source with three images. The trend of the points plotted is very similar to what we saw in A370. Other sources for Irtysh III also exhibit analogous trends. The similarity of these results for Irtysh III with A370 certainly implies that the shifting of the back projected images and caustics is proportional to the mass ratio of upper and lower extra mass clumps. 

Thus, we are reasonably confident in concluding that there exist outlying masses, well outside to the north and the south of the strong lensing region in A370. It is possibly a filament-like large scale structure passing through the cluster of galaxies. We expected to see this filament in weak lensing reconstructions, but \citet{Strait2018}, who use both strong and weak lensing data, do not seem to have it in their mass maps. However, their field of view of $\sim\!1$ Mpc may be too small to detect a filament.

\subsection{Probabilities of Magnification in the Source Plane}
\label{sec:magprob}

Galaxy clusters are often used as nature's telescopes to magnify very high redshift sources \citep{Bouwens2017,Livermore2017,Atek2018, Ishigaki2018}. The ultimate goal is to estimate the source luminosity function, but an intermediate step, which we carry out in this subsection, is to obtain a statistical characterization of a cluster's magnifying power, i.e., a probability distribution of source plane magnifications. 

The probability of magnification in the source plane can be defined as the fraction of the area in the source plane at a given redshift, $z_{\rm s}$, that has an unsigned magnification value larger than a given value $\mu$ \citep{Wong2012,Diego2019}. We denote this quantity by $\sigma_{\rm sp}(\mu,z_{\rm s})$. Our definition of $\sigma_{\rm sp}(\mu,z_{\rm s})$ differs from the more common definition of the quantity where physical source plane area is considered. $\sigma_{\rm sp}(\mu,z_{\rm s})$ will be later used to estimate the number of galaxies expected to be observed above a given detection threshold, i.e., the lensed luminosity function (see Section~\ref{sec:LF}). 

Near a caustic, this probability is proportional to the probability of a source point to be closer than a distance $\Delta \beta$ from the caustic. This relation can be approximated to $\sigma_{\rm sp} \propto \sqrt{\Delta \beta^{-1}} \propto \mu^{-2}$, \citep[see chapter 11 of][]{Schneider1992}. Thus, for high magnifications, $\sigma_{\rm sp}$ vs. $\mu$ in the log-log space can be approximated by a straight line with a slope of -2.

\begin{figure}
	\includegraphics[width=\columnwidth]{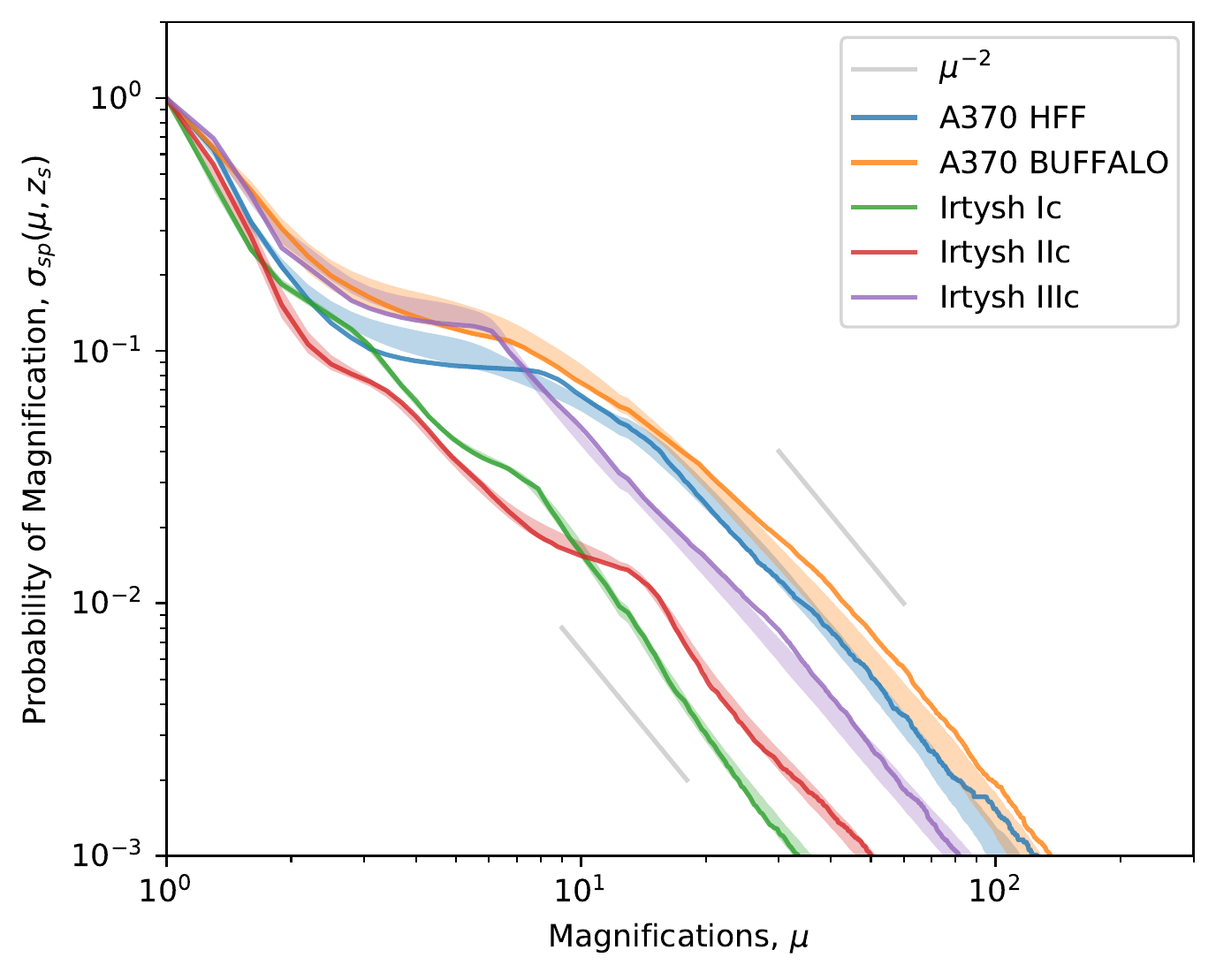}
    \caption{Cumulative probabilities of magnification,  $\sigma_{\rm sp}(\mu,z_{\rm s})$, as a function of magnifications ($\mu$) for A370 HFF and BUFFALO reconstructions and Irtysh c reconstructions (with $\sim 150$ images), for sources at $z_{\rm s}=9.0$. The colored shaded regions are showing the uncertainties in the probabilities (68 per cent confidence range) for the corresponding cluster. The gray lines denote a slope of $\mu^{-2}$. The $\mu^{-2}$ behavior of the probabilities are evident for higher magnification values. The normalization of the curves are arbitrary for different clusters given that the source plane area we have considered are different for each of them.}
    \label{fig:mf}
\end{figure}

Using the best fit \grale reconstruction we computed the magnification maps in the source plane, as described in Section~\ref{sec:magsp}, at a redshift of $z_{\rm s}=9.0$. From these magnification maps we estimated the probability of magnification, $\sigma_{\rm sp}$, for the given source plane area. Figure~\ref{fig:mf} shows $\sigma_{\rm sp}$ as a function of magnifications ($\mu$), in log-log space, for different reconstructions of A370 (with HFFv4 and BUFFALO data) and Irtysh (I, II and III; c reconstructions only, with $151$ images). The best fit distributions are shown by solid colored lines and the shaded colored regions show the uncertainties in the probabilities obtained from 40 different \grale runs. The gray lines in the figure represents the slope of $-2$. It can be seen that, for magnifications above $10$, the expected $\mu^{-2}$ behavior is fairly maintained in all of the reconstructions.

At lower magnifications,  the shape of the magnification probability distribution depends on the details of the mass distribution. In Figure~\ref{fig:mf} we see flat plateaus in the probability curves between the magnification values 2 and 10 for both A370 and Irtysh IIIc, whereas for Irtysh I and II, the $\mu^{-2}$ behavior is maintained even at lower magnifications. We speculate that these are due to the presence of naked cusps and diamond caustic folds that are outside the oval caustic in the source plane for A370 and Irtysh IIIc, which, in turn, are mostly due to the masses above and below the main cluster (see Figure~\ref{fig:A370magmaps} and \ref{fig:irtyshiiimag}). Near these regions the magnification values tend to change very rapidly. In contrast, Irtysh I and II have no such naked cusps (see Figure~\ref{fig:irtyshimagmaps} for Irtysh I) and thus have less abrupt changes in slope values of the probabilities for low magnifications.

\subsection{Redshift Dependence of the Magnification Probabilities}
\label{sec:magz}

\begin{figure}
	\includegraphics[width=\columnwidth]{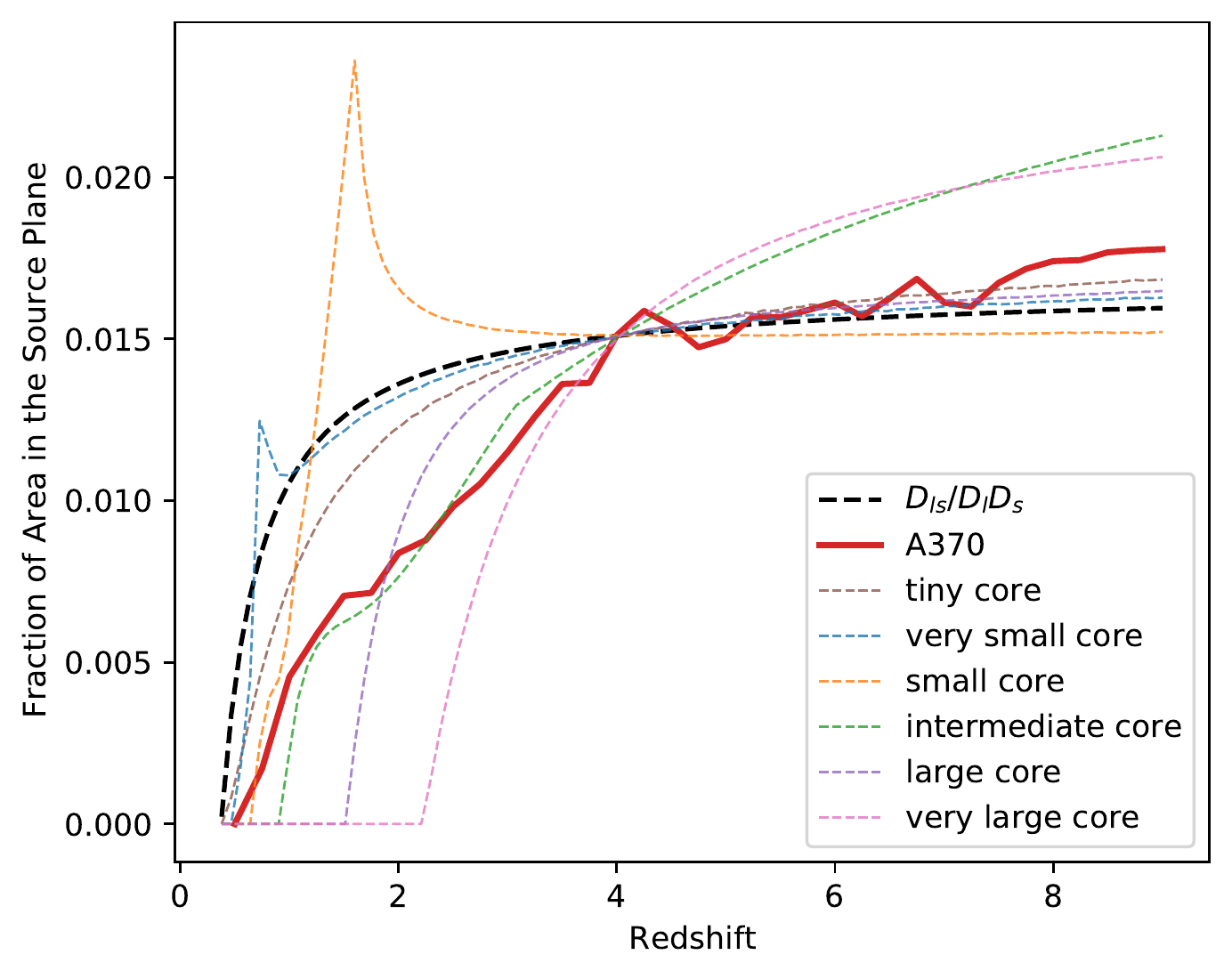}
    \caption{Redshift dependence of the probabilities of magnification in the source plane. The red line denotes the fractional area in the source plane with magnifications greater than 30, $\sigma_{\rm sp}(\mu=30,z_{\rm s})$ for A370. To reduce the computational time a coarse source plane grid was chosen for this exercise, which in turn produced a less smoother curve. The black dashed line shows the relation $\theta_E^2\!\propto\! D_{\rm ls}/(D_{\rm l}D_{\rm s})$, normalized by $\sigma_{\rm sp}(\mu=30,z_{\rm s}=4)$ of A370. The colored dotted lines show the source plane redshift dependence of $\sigma_{\rm sp}(\mu=30,z_{\rm s})$ for a six single circularly symmetric lens with varying core sizes, normalized by $\sigma_{\rm sp}(\mu=30,z_{\rm s}=4)$ of A370.}
    \label{fig:magzdepenA370}
\end{figure}

Here we continue to examine how the magnification properties of a cluster are related to those of its mass distribution. In the previous subsection we used a fixed $z_{\rm s}=9.0$, and varied magnification $\mu$. Here we fix $\mu$, and vary $z_{\rm s}$.

For a given $z_{\rm l}$, the area in the source plane, $\sigma_{\rm sp}(\mu,z_{\rm s})$, where sources at a redshift $z_{\rm s}$ must lie in order to be magnified by greater then a certain unsigned value, $\mu$, is a function of the source redshift and the amount and nature of substructure, or clumpiness in the lens. This was nicely highlighted by \cite{Vega2019}. The naive dependence on $z_{\rm s}$, which ignores clumpiness, scales as the solid angle subtended by a disk of critical density with the Einstein radius, and is given by $\theta_{\rm E}^2\propto D_{\rm ls}/(D_{\rm l}D_{\rm s})$. It is shown as the black dashed line in Figure~\ref{fig:magzdepenA370}.  (Note that the absolute normalization of the curve is not important.)

The actual dependence on $z_{\rm s}$ is more complicated because of the clumpiness in the lens; we will examine it here, in a simplified form. We consider a single circularly symmetric lens, i.e., a substructure, with a range of core radii (six dotted curves in Figure~\ref{fig:magzdepenA370}). A realistic lens will have many such substructures. 

There are two ways that a given mass distribution can attain high magnifications, say $\mu\!>\!30$. The lens must either be super-critical, or sub-critical, with $1-\mu^{-1/2}\!\lesssim\!\kappa\!<\!1$. 

A lens with a small enough core will be super-critical even at low $z_{\rm s}$. An example is the orange dotted line in Figure~\ref{fig:magzdepenA370}. There is a small flattening of the curve at $z_{\rm s}\approx 0.8$, where the lens becomes super-critical. At higher $z_{\rm s}$ it has two critical curves, radial and tangential, so there are two annular regions in the lens plane with $\mu\!>\!30$. As $z_{\rm s}$ increases, the lens plane region between them drops to lower magnifications. The source redshift where it just falls below $\mu=30$ corresponds to the location of the spike in Figure~\ref{fig:magzdepenA370}; for the orange curve that happens at $z_{\rm s}\approx 1.5$. These two features---transition to super-critical and behavior of the region between the two critical curves---are also visible in other dotted lines even though for some they are barely visible, like in the green curve at $z_{\rm s}\approx 3.0$. 

The set of six dotted curves---from the tiny to very large core---shows the full range of behavior; all normalized to go through the same $\sigma_{\rm sp}$ at $z_{\rm s}=4.0$.

A cluster mass distribution can be approximately thought of as a superposition of many power law lenses with varying normalizations, and core sizes. Equivalently, one can think of power spectrum of projected density fluctuations \citep{Mohammed2016}. If singular, or small-core substructures dominate, the overall $\sigma_{\rm sp}$ for a cluster lens will follow $\theta_E^2(z_{\rm s})$, especially at higher $z_{\rm s}$. This is the case for most of the cluster models presented in \cite{Vega2019}: their substructures (i.e., galaxies) tend to be sub-critical for $z_{\rm s}\lesssim 2$, and super-critical for sources at higher redshifts. 

Our reconstruction of A370 is represented by the thick solid red line in Figure~\ref{fig:magzdepenA370}, and shows a somewhat different behavior. It grows slower with $z_{\rm s}$ compared to models in \cite{Vega2019}, and continues to grow, though at a slower rate even past $z_{\rm s}\sim 4.0$. This is consistent with the visual impression of \grale\!'s reconstructions compared to that of many other methods: its substructure is more extended, or diffuse. Furthermore, the fact that its $\sigma_{\rm sp}$ continues to grow for all $z_{\rm s}$ means that the density fluctuation power spectrum has power on a much wider range of scales compared to that of other lens inversion methods. This conclusion is consistent with the analysis of \cite{Mohammed2016}; see their figure 9. We note that A370 has a much more complicated structure compared to the simplified analysis of this section, but it should not affect the overall behaviour.

The conclusion drawn from our Figure~\ref{fig:magzdepenA370} and figure 8 of \cite{Vega2019} is that the nature of mass clumpiness in the cluster can lead to $\sim 10-20\%$ variation in the predicted space number density of highly magnified sources at high redshifts.

\section{Effect of Lensing on the Luminosity function}
\label{sec:LF}

Working as natural telescopes, clusters of galaxies can lens distant background galaxies magnifying their size and apparent brightness. Thus otherwise unobserved galaxies come within the detection threshold of the contemporary telescopes, like the \hst, by virtue of the magnification boost provided by the gravitational lenses. 

\begin{figure}
	\includegraphics[width=\columnwidth]{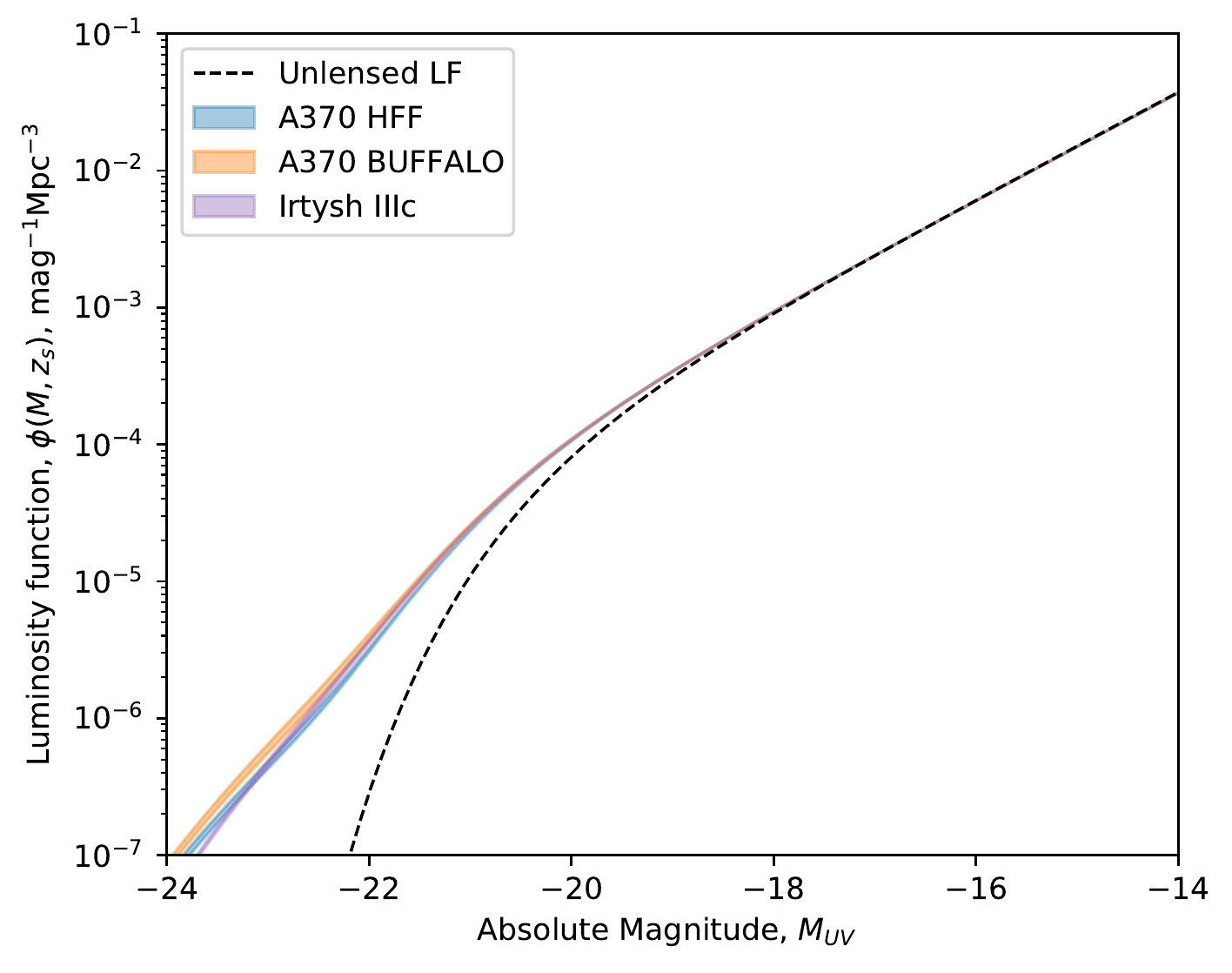}
    \caption{Luminosity functions at $z_{\rm s}=9.0$. The colored shaded regions represent the uncertainties (68 per cent confidence range) in the lensed luminosity functions for the \grale reconstructions of A370 (HFF and BUFFALO models) and of Irtysh IIIc reconstruction. The black dashed line represents the classical unlensed Schechter function at a redshift of $z_{\rm s}=9.0$ and $\Delta z_{\rm s}=1.0$. }
    \label{fig:lfirtysh}
\end{figure}

The luminosity distribution for the background distant galaxies are described by the  luminosity  functions (LF). In this work, we are using the classical Schechter LF \citep{Schechter1976}, which for a given redshift $z_{\rm s}$ can be written in terms of absolute magnitude, M, as,
\begin{equation}
    \phi(M,z_{\rm s})=0.4 \ln (10) \phi^* 10^{-0.4 (M-M_*)(\alpha+1)} {\rm e}^{-10^{-0.4 (M-M_*)}},
    \label{eq:sf}
\end{equation}
where the values of normalization constant $\phi^*$, characteristic magnitude $M_*$ and faint-end logarithmic slope $\alpha$ are the best-fitting parameters at the given redshift. In our work their values are taken from \citet{Ishigaki2018}.
The use of the Schechter LF at high redshifts is further justified by \cite{Bouwens2015}.
The UV LF is shown as the black dashed line in Figure~\ref{fig:lfirtysh}, for redshift $z_{\rm s}=9.0$, with $\Delta z_{\rm s}=1.0$.

When distant galaxies with a given intrinsic luminosity function get lensed by a cluster, their flux is magnified by a factor of $\mu$ and they gain an apparent magnitude of $-2.5 \log_{10} \mu$. On the other hand, due to lensing the effective sampling volume reduces by a factor of $1/\mu$ \citep{Broadhurst1995}. One can compute the lensed luminosity functions $\phi'(M,z_{\rm s})$ by convolving the known unlensed luminosity functions $\phi(M,z_{\rm s})$ with the distribution of the source plane magnifications at a given redshift of $z_{\rm s}$ and a given redshift interval of $\Delta z_{\rm s}$,
\begin{equation}
    \phi'(M,z_{\rm s})=\int_{\mu_{\rm min}}^{\mu_{\rm max}} \phi(M+ 2.5 \log_{10} \mu,z_{\rm s}) \frac{1}{\mu} \frac{{\rm d} \sigma_{\rm sp} (\mu, z_{\rm s})}{{\rm d} \mu} {\rm d}\mu,
    \label{eq:lf}
\end{equation}
where ${\rm d} \sigma_{\rm sp} (\mu, z_{\rm s})/{\rm d} \mu$ is the area in the source plane with magnifications between $\mu$ and $\mu+{\rm d}\mu$. Following the approach taken by \citet{Vega2019} the integral is computed in a range of $\mu_{\rm min}=1$ to $\mu_{\rm max}=100$. Demagnification is neglected since the field of view is limited to the regions where $\mu>1$. While doing the integral in Equation~\ref{eq:lf} we did not assume any magnitude cut off at either end of $\phi(M)$.

The shaded bands in Figure~\ref{fig:lfirtysh} show the lensed luminosity functions with uncertainties for the range of 40 reconstructions of A370 (HFFv4 and BUFFALO) and Irtysh IIIc. For example, at $ {\rm M}=-22.5$ the fractional uncertainty in lensed luminosity functions is $0.078$ for our A370 reconstruction with BUFFALO data; whereas it is $0.052$ for the HFF reconstruction of A370 and $0.037$ for Irtysh IIIc. All these values are at 68 per cent confidence range. The lensing affects the number density of the observed galaxies at a given absolute magnitude value. Some of the otherwise unobserved distant faint galaxies are pushed above the detection threshold due to the magnification and the already observed galaxies experience a flux increase and thus shifted towards a lower absolute magnitude. The change in the number of detectable galaxies depends on the faint-end logarithmic slope, $\alpha$. A value of $\alpha=-2$ implies a magnification-independent luminosity function \citep{Broadhurst1995}. In this work we are using, $\alpha=-1.98$, as estimated by \citet{Ishigaki2018} for $z_{\rm s}=9.0$. Since in our case $\alpha>-2$ there will be a slight deficit of galaxies at fainter magnitudes.  On the other hand at brighter magnitudes there is an excess of galaxies due to lensing. These explain why we see a deviation of the lensed LF from the unlensed LF in Figure~\ref{fig:lfirtysh}. These results are consistent with other works, for example, see section 4 of \citet{Vega2019}.

\section{Conclusions}

Using the strong lensing image catalogs from the BUFFALO collaboration we have performed the reconstruction of the galaxy cluster Abell 370 ($z_{\rm l}=0.375$) with our free-form algorithm \textsc{grale}. The lens plane rms is $0.45$ arcsec, significantly lower than our HFF v4 reconstruction of the same cluster, which was $0.88$ arcsec. Since the number of lensed images is nearly the same in both inversions, the improved LPrms is an indication of the improvement in the data quality of BUFFALO.

The reconstruction confirms the previously established active merging state of the cluster, as indicated by two large mass peaks separated by $\sim 200$  kpc along the N-S direction.  The fact that in our reconstructions the northern mass peak is displaced from the observed light peak of the northern BCG, could be a further indication of the disturbed state of the cluster. Though details vary, this displacement is also present in almost all the publicly available reconstructions of A370 (see Table~\ref{tab:3massfeatures}), using parametric, free-form, and hybrid methods. Because this feature appears to be robust, it could provide  another argument in favor of particle nature of dark matter, which uses baryonic matter as the source and cannot easily explain offsets between mass and light. 

Our mass map shows two other interesting features. One, is a roughly critical density mass clump, of $M\sim 3\times 10^{12} M_{\sun}$, spanning $\sim100$ kpc, in the eastern side of the cluster (see Section~\ref{sec:substructure}). It was recovered independently in HFF reconstructions by free-form \grale, hybrid \textsc{wslap+}, and is present in some parametric models as a cluster-scale dark matter halo. Our reconstruction using BUFFALO data also recovered this mass clump, further supporting its possibility. 

The other mass feature is a probable filament (or similar external mass) stretching N-S through the cluster, and extending well outside the strong lensing region (see Section~\ref{sec:filament}). Some of the publicly available reconstructions of A370 by other groups include similar features:
external shear, or highly elongated mass `fingers' going N-S through the cluster center. Since our modeling window extends only somewhat beyond the strong lensing region, to account for the external mass \grale generates mass clumps along the same direction, towards the lesser-constrained (containing few or no images) edges of the reconstruction window. 
To verify this hypothesis we created the synthetic cluster, Irtysh III (see Section~\ref{sec:irtyshiii}), 
with two external mass clumps, mimicking a filament. 
The recovered mass maps of Irtysh III are consistent with our expectations: \grale produced mass clumps near the north and south edges of the modeling  region, very similar to those in A370. (For Irtysh I, the synthetic cluster used in G20, where no external masses were included, \grale did not produce this kind of extra N-S mass structures.)

Since the external mass responsible for the N-S mass clumps is well outside the multiple image region, its exact nature is not well constrained. It could be another galaxy cluster to the north or south of A370, or a large scale structure filament, with mass distributed roughly evenly between north and south. Our reconstructions favor the latter, because individual \grale runs place varying amounts of mass at the northern vs. southern edges: there is no strong preference for one over the other.  The galaxy distribution identified by \cite{Lagattuta2019} from the CFHT data, and extending NNW and SSE of the strong lensing region to the virial radius of the cluster, may be part of this potential filament structure. 

While it is difficult to identify filaments in either ground-based or space-based optical or weak lensing data, our analysis suggests that it might be possible that the strong lensing data, used as part of parametric or free-form methods, has the potential to help identify filaments.

 In Section~\ref{sec:magprob} and \ref{sec:magz} we consider the magnification properties of A370, as reconstructed by \grale\!\!, and also relate the specific features of the magnification probability distribution, and magnification's dependence on source redshift on the properties of the mass distribution. In Section~\ref{sec:magprob},
using  a very accurate forward lensing method, we calculate the source plane magnification distributions of Abell 370, and estimate the probabilities of magnifications, $\sigma_{\rm sp}(\mu,z_{\rm s})$, i.e., the fraction of source plane area above a certain magnification value, at $z_{\rm s}=9.0$. We found that all the reconstructions exhibit the expected $\sigma_{\rm sp}(\mu,z_{\rm s}=9.0)\propto\mu^{-2}$ behavior at large magnifications ($\mu>10$). At lower magnifications, the behavior of $\sigma_{\rm sp}$ depends on the specifics of the recovered mass distribution. It differs slightly between the reconstructions, depending on whether the diamond caustics are enclosed by, or protrude outside of the oval caustic.
We also explored the redshift dependence of $\sigma_{\rm sp}(\mu,z_{\rm s})$ for a given magnification value of $\mu=30$ (Section~\ref{sec:magz}). When compared to other models, our magnification probabilities $\sigma_{\rm sp}(\mu,z_{\rm s})$ tend to grow slower with redshift, implying that the substructure in \grale reconstructions spans a wider range of scales compared to that of other reconstructions. 

Finally, in Section~\ref{sec:LF}, we estimated the uncertainties in the lensed luminosity functions for our A370 reconstruction at a redshift of $z_{\rm s}=9.0$.
We found the effect of lensing on the luminosity functions with our reconstruction method to be in agreement with concurrent works by other groups.

One of the main conclusions from this work is based on the three mass features discussed above. We find it interesting, but do not fully understand why many reconstructions with distinct modelling philosophies -  parametric vs. free-form - can lead to converging results regarding specific mass features, for example the $\sim 100$ kpc mass east of cluster center, while at the same time, models of the same type using the same algorithm can draw contrasting conclusions, just by using slightly varied model priors and data constraints. Despite of this, we suggest that mass features that are common across the majority of different reconstructions - like at least two of the features considered here - can be considered as potential indicators of real structures.

We further conclude that because free-form models enjoy the freedom from fixed parameter spaces of parametric models, it enables them to recover mass features that are elusive for some parametric models, especially if the quantity and quality of lensing constraints is sub-optimal.  All three mass features discussed in Section~\ref{sec:recres} were present in free-form reconstructions since the very first model by \citet{Abdelsalam1998}, two decades ago, and all \grale reconstructions, since HFFv1. But this freedom of free-form methods comes at a cost of lower spatial resolutions and possible overestimation of uncertainties. Free-form methods are also susceptible to producing artefacts when image number density is low. With upcoming observational facilities like the \textit{James Webb Space Telescope}, which is expected to uncover hundreds of new strongly lensed images, free-form methods will be well positioned to accurately map out cluster mass distribution on scales greater than most individual galaxies.

\section*{Acknowledgements}

The authors would like to thank the anonymous reviewer for their helpful comments that improved the paper. 
AG and LLRW acknowledge the support and the computational resources of Minnesota Supercomputing Institute (MSI) which were crucial for this work. MJ is supported by the United Kingdom Research and Innovation (UKRI) Future Leaders Fellowship `Using Cosmic Beasts to uncover the Nature of Dark Matter' (grant number MR/S017216/1). GM received funding from the European Union’s Horizon 2020 research and innovation programme under the Marie Skłodowska-Curie grant agreement No MARACAS - DLV-896778.

% \vspace{-0.5cm}

\section*{Data Availability}

The lens inversion data underlying this article will be shared on reasonable request to the corresponding author. The strong lensing image set used in this work (see Table~\ref{tab:imgdata}) is available as online supplementary material to this article.

%%%%%%%%%%%%%%%%%%%%%%%%%%%%%%%%%%%%%%%%%%%%%%%%%%



\bibliographystyle{mnras}
\bibliography{ref}

\begin{thebibliography}{}
\makeatletter
\relax
\def\mn@urlcharsother{\let\do\@makeother \do\$\do\&\do\#\do\^\do\_\do\%\do\~}
\def\mn@doi{\begingroup\mn@urlcharsother \@ifnextchar [ {\mn@doi@}
  {\mn@doi@[]}}
\def\mn@doi@[#1]#2{\def\@tempa{#1}\ifx\@tempa\@empty \href
  {http://dx.doi.org/#2} {doi:#2}\else \href {http://dx.doi.org/#2} {#1}\fi
  \endgroup}
\def\mn@eprint#1#2{\mn@eprint@#1:#2::\@nil}
\def\mn@eprint@arXiv#1{\href {http://arxiv.org/abs/#1} {{\tt arXiv:#1}}}
\def\mn@eprint@dblp#1{\href {http://dblp.uni-trier.de/rec/bibtex/#1.xml}
  {dblp:#1}}
\def\mn@eprint@#1:#2:#3:#4\@nil{\def\@tempa {#1}\def\@tempb {#2}\def\@tempc
  {#3}\ifx \@tempc \@empty \let \@tempc \@tempb \let \@tempb \@tempa \fi \ifx
  \@tempb \@empty \def\@tempb {arXiv}\fi \@ifundefined
  {mn@eprint@\@tempb}{\@tempb:\@tempc}{\expandafter \expandafter \csname
  mn@eprint@\@tempb\endcsname \expandafter{\@tempc}}}

\bibitem[\protect\citeauthoryear{{Abdelsalam}, {Saha}  \&
  {Williams}}{{Abdelsalam} et~al.}{1998}]{Abdelsalam1998}
{Abdelsalam} H.~M.,  {Saha} P.,   {Williams} L. L.~R.,  1998, \mn@doi [\mnras]
  {10.1046/j.1365-8711.1998.01356.x}, \href
  {https://ui.adsabs.harvard.edu/abs/1998MNRAS.294..734A} {294, 734}

\bibitem[\protect\citeauthoryear{{Atek}, {Richard}, {Kneib}  \&
  {Schaerer}}{{Atek} et~al.}{2018}]{Atek2018}
{Atek} H.,  {Richard} J.,  {Kneib} J.-P.,   {Schaerer} D.,  2018, \mn@doi
  [\mnras] {10.1093/mnras/sty1820}, \href
  {https://ui.adsabs.harvard.edu/abs/2018MNRAS.479.5184A} {479, 5184}

\bibitem[\protect\citeauthoryear{{Bahcall} \& {Cen}}{{Bahcall} \&
  {Cen}}{1993}]{Bahcall1993}
{Bahcall} N.~A.,  {Cen} R.,  1993, \mn@doi [\apjl] {10.1086/186803}, \href
  {https://ui.adsabs.harvard.edu/abs/1993ApJ...407L..49B} {407, L49}

\bibitem[\protect\citeauthoryear{{Bartelmann}}{{Bartelmann}}{2010}]{Bartelmann2010}
{Bartelmann} M.,  2010, \mn@doi [Classical and Quantum Gravity]
  {10.1088/0264-9381/27/23/233001}, \href
  {https://ui.adsabs.harvard.edu/abs/2010CQGra..27w3001B} {27, 233001}

\bibitem[\protect\citeauthoryear{{Bouwens} et~al.,}{{Bouwens}
  et~al.}{2015}]{Bouwens2015}
{Bouwens} R.~J.,  et~al., 2015, \mn@doi [\apj] {10.1088/0004-637X/803/1/34},
  \href {https://ui.adsabs.harvard.edu/abs/2015ApJ...803...34B} {803, 34}

\bibitem[\protect\citeauthoryear{{Bouwens}, {Illingworth}, {Oesch}, {Atek},
  {Lam}  \& {Stefanon}}{{Bouwens} et~al.}{2017a}]{Bouwens2017a}
{Bouwens} R.~J.,  {Illingworth} G.~D.,  {Oesch} P.~A.,  {Atek} H.,  {Lam} D.,
  {Stefanon} M.,  2017a, \mn@doi [\apj] {10.3847/1538-4357/aa74e4}, \href
  {https://ui.adsabs.harvard.edu/abs/2017ApJ...843...41B} {843, 41}

\bibitem[\protect\citeauthoryear{{Bouwens}, {Oesch}, {Illingworth}, {Ellis}  \&
  {Stefanon}}{{Bouwens} et~al.}{2017b}]{Bouwens2017}
{Bouwens} R.~J.,  {Oesch} P.~A.,  {Illingworth} G.~D.,  {Ellis} R.~S.,
  {Stefanon} M.,  2017b, \mn@doi [\apj] {10.3847/1538-4357/aa70a4}, \href
  {https://ui.adsabs.harvard.edu/abs/2017ApJ...843..129B} {843, 129}

\bibitem[\protect\citeauthoryear{{Broadhurst}, {Taylor}  \&
  {Peacock}}{{Broadhurst} et~al.}{1995}]{Broadhurst1995}
{Broadhurst} T.~J.,  {Taylor} A.~N.,   {Peacock} J.~A.,  1995, \mn@doi [\apj]
  {10.1086/175053}, \href
  {https://ui.adsabs.harvard.edu/abs/1995ApJ...438...49B} {438, 49}

\bibitem[\protect\citeauthoryear{{Diego}}{{Diego}}{2019}]{Diego2019}
{Diego} J.~M.,  2019, \mn@doi [\aap] {10.1051/0004-6361/201833670}, \href
  {https://ui.adsabs.harvard.edu/abs/2019A&A...625A..84D} {625, A84}

\bibitem[\protect\citeauthoryear{{Diego} et~al.,}{{Diego}
  et~al.}{2018}]{Diego2018b}
{Diego} J.~M.,  et~al., 2018, \mn@doi [\mnras] {10.1093/mnras/stx2609}, \href
  {https://ui.adsabs.harvard.edu/abs/2018MNRAS.473.4279D} {473, 4279}

\bibitem[\protect\citeauthoryear{{Faber} \& {Jackson}}{{Faber} \&
  {Jackson}}{1976}]{Faber1976}
{Faber} S.~M.,  {Jackson} R.~E.,  1976, \mn@doi [\apj] {10.1086/154215}, \href
  {https://ui.adsabs.harvard.edu/abs/1976ApJ...204..668F} {204, 668}

\bibitem[\protect\citeauthoryear{{Fort}, {Mellier}, {Picat}, {Rio}  \&
  {Lelievre}}{{Fort} et~al.}{1986}]{Fort1986}
{Fort} B.,  {Mellier} Y.,  {Picat} J.~P.,  {Rio} Y.,   {Lelievre} G.,  1986, in
  {Crawford} D.~L.,  ed.,  Society of Photo-Optical Instrumentation Engineers
  (SPIE) Conference Series Vol. 627, Instrumentation in astronomy VI. pp
  321--327, \mn@doi{10.1117/12.968105}

\bibitem[\protect\citeauthoryear{{Ghosh}, {Williams}  \& {Liesenborgs}}{{Ghosh}
  et~al.}{2020}]{Ghosh2020}
{Ghosh} A.,  {Williams} L. L.~R.,   {Liesenborgs} J.,  2020, \mn@doi [\mnras]
  {10.1093/mnras/staa962}, \href
  {https://ui.adsabs.harvard.edu/abs/2020MNRAS.494.3998G} {494, 3998}

\bibitem[\protect\citeauthoryear{{Hammer}}{{Hammer}}{1987}]{Hammer1987}
{Hammer} F.,  1987, in High Redshift and Primeval Galaxies. pp 467--473

\bibitem[\protect\citeauthoryear{{Hammer} \& {Rigaut}}{{Hammer} \&
  {Rigaut}}{1989}]{Hammer1989}
{Hammer} F.,  {Rigaut} F.,  1989, \aap, \href
  {https://ui.adsabs.harvard.edu/abs/1989A&A...226...45H} {226, 45}

\bibitem[\protect\citeauthoryear{{Hoekstra}}{{Hoekstra}}{2007}]{Hoekstra2007}
{Hoekstra} H.,  2007, \mn@doi [\mnras] {10.1111/j.1365-2966.2007.11951.x},
  \href {https://ui.adsabs.harvard.edu/abs/2007MNRAS.379..317H} {379, 317}

\bibitem[\protect\citeauthoryear{{Ishigaki}, {Kawamata}, {Ouchi}, {Oguri},
  {Shimasaku}  \& {Ono}}{{Ishigaki} et~al.}{2018}]{Ishigaki2018}
{Ishigaki} M.,  {Kawamata} R.,  {Ouchi} M.,  {Oguri} M.,  {Shimasaku} K.,
  {Ono} Y.,  2018, \mn@doi [\apj] {10.3847/1538-4357/aaa544}, \href
  {https://ui.adsabs.harvard.edu/abs/2018ApJ...854...73I} {854, 73}

\bibitem[\protect\citeauthoryear{{Jauzac} et~al.,}{{Jauzac}
  et~al.}{2015}]{Jauzac2015}
{Jauzac} M.,  et~al., 2015, \mn@doi [\mnras] {10.1093/mnras/stv1402}, \href
  {https://ui.adsabs.harvard.edu/abs/2015MNRAS.452.1437J} {452, 1437}

\bibitem[\protect\citeauthoryear{{Johnson}, {Sharon}, {Bayliss}, {Gladders},
  {Coe}  \& {Ebeling}}{{Johnson} et~al.}{2014}]{Johnson2014}
{Johnson} T.~L.,  {Sharon} K.,  {Bayliss} M.~B.,  {Gladders} M.~D.,  {Coe} D.,
   {Ebeling} H.,  2014, \mn@doi [\apj] {10.1088/0004-637X/797/1/48}, \href
  {https://ui.adsabs.harvard.edu/abs/2014ApJ...797...48J} {797, 48}

\bibitem[\protect\citeauthoryear{{Kassiola} \& {Kovner}}{{Kassiola} \&
  {Kovner}}{1993}]{Kovner1993}
{Kassiola} A.,  {Kovner} I.,  1993, \mn@doi [\apj] {10.1086/173325}, \href
  {https://ui.adsabs.harvard.edu/abs/1993ApJ...417..450K} {417, 450}

\bibitem[\protect\citeauthoryear{{Kawamata}, {Ishigaki}, {Shimasaku}, {Oguri},
  {Ouchi}  \& {Tanigawa}}{{Kawamata} et~al.}{2018}]{Kawamata2018}
{Kawamata} R.,  {Ishigaki} M.,  {Shimasaku} K.,  {Oguri} M.,  {Ouchi} M.,
  {Tanigawa} S.,  2018, \mn@doi [\apj] {10.3847/1538-4357/aaa6cf}, \href
  {https://ui.adsabs.harvard.edu/abs/2018ApJ...855....4K} {855, 4}

\bibitem[\protect\citeauthoryear{{Keeton}}{{Keeton}}{2001}]{Keeton2001}
{Keeton} C.~R.,  2001, arXiv e-prints, \href
  {https://ui.adsabs.harvard.edu/abs/2001astro.ph..2341K} {pp
  astro--ph/0102341}

\bibitem[\protect\citeauthoryear{{Kneib} \& {Natarajan}}{{Kneib} \&
  {Natarajan}}{2011}]{Kneib2011}
{Kneib} J.-P.,  {Natarajan} P.,  2011, \mn@doi [\aapr]
  {10.1007/s00159-011-0047-3}, \href
  {https://ui.adsabs.harvard.edu/abs/2011A&ARv..19...47K} {19, 47}

\bibitem[\protect\citeauthoryear{{Kneib}, {Mellier}, {Fort}  \&
  {Mathez}}{{Kneib} et~al.}{1993}]{Kneib1993}
{Kneib} J.~P.,  {Mellier} Y.,  {Fort} B.,   {Mathez} G.,  1993, \aap, \href
  {https://ui.adsabs.harvard.edu/abs/1993A&A...273..367K} {273, 367}

\bibitem[\protect\citeauthoryear{{Kovner}}{{Kovner}}{1989}]{Kovner1989}
{Kovner} I.,  1989, \mn@doi [\apj] {10.1086/167133}, \href
  {https://ui.adsabs.harvard.edu/abs/1989ApJ...337..621K} {337, 621}

\bibitem[\protect\citeauthoryear{{Lagattuta} et~al.,}{{Lagattuta}
  et~al.}{2017}]{Lagattuta2017}
{Lagattuta} D.~J.,  et~al., 2017, \mn@doi [\mnras] {10.1093/mnras/stx1079},
  \href {https://ui.adsabs.harvard.edu/abs/2017MNRAS.469.3946L} {469, 3946}

\bibitem[\protect\citeauthoryear{{Lagattuta} et~al.,}{{Lagattuta}
  et~al.}{2019}]{Lagattuta2019}
{Lagattuta} D.~J.,  et~al., 2019, \mn@doi [\mnras] {10.1093/mnras/stz620},
  \href {https://ui.adsabs.harvard.edu/abs/2019MNRAS.485.3738L} {485, 3738}

\bibitem[\protect\citeauthoryear{{Liesenborgs} \& {De Rijcke}}{{Liesenborgs} \&
  {De Rijcke}}{2012}]{Liesenborgs2012}
{Liesenborgs} J.,  {De Rijcke} S.,  2012, \mn@doi [\mnras]
  {10.1111/j.1365-2966.2012.21751.x}, \href
  {https://ui.adsabs.harvard.edu/abs/2012MNRAS.425.1772L} {425, 1772}

\bibitem[\protect\citeauthoryear{{Liesenborgs}, {De Rijcke}  \&
  {Dejonghe}}{{Liesenborgs} et~al.}{2006}]{Liesenborgs2006a}
{Liesenborgs} J.,  {De Rijcke} S.,   {Dejonghe} H.,  2006, \mn@doi [\mnras]
  {10.1111/j.1365-2966.2006.10040.x}, \href
  {https://ui.adsabs.harvard.edu/abs/2006MNRAS.367.1209L} {367, 1209}

\bibitem[\protect\citeauthoryear{{Liesenborgs}, {de Rijcke}, {Dejonghe}  \&
  {Bekaert}}{{Liesenborgs} et~al.}{2007}]{Liesenborgs2007}
{Liesenborgs} J.,  {de Rijcke} S.,  {Dejonghe} H.,   {Bekaert} P.,  2007,
  \mn@doi [\mnras] {10.1111/j.1365-2966.2007.12236.x}, \href
  {https://ui.adsabs.harvard.edu/abs/2007MNRAS.380.1729L} {380, 1729}

\bibitem[\protect\citeauthoryear{{Liesenborgs}, {Williams}, {Wagner}  \& {De
  Rijcke}}{{Liesenborgs} et~al.}{2020}]{Liesenborgs2020}
{Liesenborgs} J.,  {Williams} L. L.~R.,  {Wagner} J.,   {De Rijcke} S.,  2020,
  \mn@doi [\mnras] {10.1093/mnras/staa842}, \href
  {https://ui.adsabs.harvard.edu/abs/2020MNRAS.494.3253L} {494, 3253}

\bibitem[\protect\citeauthoryear{{Limousin} et~al.,}{{Limousin}
  et~al.}{2016}]{Limousin2016}
{Limousin} M.,  et~al., 2016, \mn@doi [\aap] {10.1051/0004-6361/201527638},
  \href {https://ui.adsabs.harvard.edu/abs/2016A&A...588A..99L} {588, A99}

\bibitem[\protect\citeauthoryear{{Livermore}, {Finkelstein}  \&
  {Lotz}}{{Livermore} et~al.}{2017}]{Livermore2017}
{Livermore} R.~C.,  {Finkelstein} S.~L.,   {Lotz} J.~M.,  2017, \mn@doi [\apj]
  {10.3847/1538-4357/835/2/113}, \href
  {https://ui.adsabs.harvard.edu/abs/2017ApJ...835..113L} {835, 113}

\bibitem[\protect\citeauthoryear{{Lotz} et~al.,}{{Lotz}
  et~al.}{2017}]{Lotz2017}
{Lotz} J.~M.,  et~al., 2017, \mn@doi [\apj] {10.3847/1538-4357/837/1/97}, \href
  {https://ui.adsabs.harvard.edu/abs/2017ApJ...837...97L} {837, 97}

\bibitem[\protect\citeauthoryear{{Lynds} \& {Petrosian}}{{Lynds} \&
  {Petrosian}}{1989}]{Lynds1989}
{Lynds} R.,  {Petrosian} V.,  1989, \mn@doi [\apj] {10.1086/166989}, \href
  {https://ui.adsabs.harvard.edu/abs/1989ApJ...336....1L} {336, 1}

\bibitem[\protect\citeauthoryear{{Mahler} et~al.,}{{Mahler}
  et~al.}{2018}]{Mahler2018}
{Mahler} G.,  et~al., 2018, \mn@doi [\mnras] {10.1093/mnras/stx1971}, \href
  {https://ui.adsabs.harvard.edu/abs/2018MNRAS.473..663M} {473, 663}

\bibitem[\protect\citeauthoryear{{Medezinski}, {Broadhurst}, {Umetsu},
  {Ben{\'\i}tez}  \& {Taylor}}{{Medezinski} et~al.}{2011}]{Medezinski2011}
{Medezinski} E.,  {Broadhurst} T.,  {Umetsu} K.,  {Ben{\'\i}tez} N.,   {Taylor}
  A.,  2011, \mn@doi [\mnras] {10.1111/j.1365-2966.2011.18332.x}, \href
  {https://ui.adsabs.harvard.edu/abs/2011MNRAS.414.1840M} {414, 1840}

\bibitem[\protect\citeauthoryear{{Mellier}, {Soucail}, {Fort}  \&
  {Mathez}}{{Mellier} et~al.}{1988}]{Mellier1988}
{Mellier} Y.,  {Soucail} G.,  {Fort} B.,   {Mathez} G.,  1988, \aap, \href
  {https://ui.adsabs.harvard.edu/abs/1988A&A...199...13M} {199, 13}

\bibitem[\protect\citeauthoryear{{Meneghetti} et~al.,}{{Meneghetti}
  et~al.}{2017}]{Meneghetti2017}
{Meneghetti} M.,  et~al., 2017, \mn@doi [\mnras] {10.1093/mnras/stx2064}, \href
  {https://ui.adsabs.harvard.edu/abs/2017MNRAS.472.3177M} {472, 3177}

\bibitem[\protect\citeauthoryear{{Mohammed}, {Liesenborgs}, {Saha}  \&
  {Williams}}{{Mohammed} et~al.}{2014}]{Mohammed2014}
{Mohammed} I.,  {Liesenborgs} J.,  {Saha} P.,   {Williams} L.~L.~R.,  2014,
  \mn@doi [\mnras] {10.1093/mnras/stu124}, \href
  {https://ui.adsabs.harvard.edu/abs/2014MNRAS.439.2651M} {439, 2651}

\bibitem[\protect\citeauthoryear{{Mohammed}, {Saha}, {Williams}, {Liesenborgs}
  \& {Sebesta}}{{Mohammed} et~al.}{2016}]{Mohammed2016}
{Mohammed} I.,  {Saha} P.,  {Williams} L. L.~R.,  {Liesenborgs} J.,   {Sebesta}
  K.,  2016, \mn@doi [\mnras] {10.1093/mnras/stw727}, \href
  {https://ui.adsabs.harvard.edu/abs/2016MNRAS.459.1698M} {459, 1698}

\bibitem[\protect\citeauthoryear{{Natarajan} \& {Kneib}}{{Natarajan} \&
  {Kneib}}{1997}]{Natarajan1997}
{Natarajan} P.,  {Kneib} J.-P.,  1997, \mn@doi [\mnras]
  {10.1093/mnras/287.4.833}, \href
  {https://ui.adsabs.harvard.edu/abs/1997MNRAS.287..833N} {287, 833}

\bibitem[\protect\citeauthoryear{{Navarro}, {Frenk}  \& {White}}{{Navarro}
  et~al.}{1996}]{Navarro1996}
{Navarro} J.~F.,  {Frenk} C.~S.,   {White} S. D.~M.,  1996, \mn@doi [\apj]
  {10.1086/177173}, \href
  {https://ui.adsabs.harvard.edu/abs/1996ApJ...462..563N} {462, 563}

\bibitem[\protect\citeauthoryear{{Press} \& {Schechter}}{{Press} \&
  {Schechter}}{1974}]{Schechter1974}
{Press} W.~H.,  {Schechter} P.,  1974, \mn@doi [\apj] {10.1086/152650}, \href
  {https://ui.adsabs.harvard.edu/abs/1974ApJ...187..425P} {187, 425}

\bibitem[\protect\citeauthoryear{{Priewe}, {Williams}, {Liesenborgs}, {Coe}  \&
  {Rodney}}{{Priewe} et~al.}{2017}]{Priewe2017}
{Priewe} J.,  {Williams} L. L.~R.,  {Liesenborgs} J.,  {Coe} D.,   {Rodney}
  S.~A.,  2017, \mn@doi [\mnras] {10.1093/mnras/stw2785}, \href
  {https://ui.adsabs.harvard.edu/abs/2017MNRAS.465.1030P} {465, 1030}

\bibitem[\protect\citeauthoryear{{Raney}, {Keeton}  \& {Brennan}}{{Raney}
  et~al.}{2020a}]{Raney2020}
{Raney} C.~A.,  {Keeton} C.~R.,   {Brennan} S.,  2020a, \mn@doi [\mnras]
  {10.1093/mnras/stz3116}, \href
  {https://ui.adsabs.harvard.edu/abs/2020MNRAS.492..503R} {492, 503}

\bibitem[\protect\citeauthoryear{{Raney}, {Keeton}, {Brennan}  \&
  {Fan}}{{Raney} et~al.}{2020b}]{Raney2020a}
{Raney} C.~A.,  {Keeton} C.~R.,  {Brennan} S.,   {Fan} H.,  2020b, \mn@doi
  [\mnras] {10.1093/mnras/staa921}, \href
  {https://ui.adsabs.harvard.edu/abs/2020MNRAS.494.4771R} {494, 4771}

\bibitem[\protect\citeauthoryear{{Remolina Gonz{\'a}lez}, {Sharon}  \&
  {Mahler}}{{Remolina Gonz{\'a}lez} et~al.}{2018}]{Gonzalez2018}
{Remolina Gonz{\'a}lez} J.~D.,  {Sharon} K.,   {Mahler} G.,  2018, \mn@doi
  [\apj] {10.3847/1538-4357/aacf8e}, \href
  {https://ui.adsabs.harvard.edu/abs/2018ApJ...863...60R} {863, 60}

\bibitem[\protect\citeauthoryear{{Richard}, {Kneib}, {Limousin}, {Edge}  \&
  {Jullo}}{{Richard} et~al.}{2010}]{Richard2010}
{Richard} J.,  {Kneib} J.~P.,  {Limousin} M.,  {Edge} A.,   {Jullo} E.,  2010,
  \mn@doi [\mnras] {10.1111/j.1745-3933.2009.00796.x}, \href
  {https://ui.adsabs.harvard.edu/abs/2010MNRAS.402L..44R} {402, L44}

\bibitem[\protect\citeauthoryear{{Richard} et~al.,}{{Richard}
  et~al.}{2014}]{Richard2014}
{Richard} J.,  et~al., 2014, \mn@doi [\mnras] {10.1093/mnras/stu1395}, \href
  {https://ui.adsabs.harvard.edu/abs/2014MNRAS.444..268R} {444, 268}

\bibitem[\protect\citeauthoryear{{Richard} et~al.,}{{Richard}
  et~al.}{2021}]{Richard2021}
{Richard} J.,  et~al., 2021, \mn@doi [\aap] {10.1051/0004-6361/202039462},
  \href {https://ui.adsabs.harvard.edu/abs/2021A&A...646A..83R} {646, A83}

\bibitem[\protect\citeauthoryear{{Saha}}{{Saha}}{2000}]{Saha2000}
{Saha} P.,  2000, \mn@doi [\aj] {10.1086/301581}, \href
  {https://ui.adsabs.harvard.edu/abs/2000AJ....120.1654S} {120, 1654}

\bibitem[\protect\citeauthoryear{{Schechter}}{{Schechter}}{1976}]{Schechter1976}
{Schechter} P.,  1976, \mn@doi [\apj] {10.1086/154079}, \href
  {https://ui.adsabs.harvard.edu/abs/1976ApJ...203..297S} {203, 297}

\bibitem[\protect\citeauthoryear{{Schneider}, {Ehlers}  \& {Falco}}{{Schneider}
  et~al.}{1992}]{Schneider1992}
{Schneider} P.,  {Ehlers} J.,   {Falco} E.~E.,  1992, {Gravitational Lenses}.
Springer-Verlag, \mn@doi{10.1007/978-3-662-03758-4}

\bibitem[\protect\citeauthoryear{{Sebesta}, {Williams}, {Mohammed}, {Saha}  \&
  {Liesenborgs}}{{Sebesta} et~al.}{2016}]{sebesta2016}
{Sebesta} K.,  {Williams} L.~L.~R.,  {Mohammed} I.,  {Saha} P.,   {Liesenborgs}
  J.,  2016, \mn@doi [\mnras] {10.1093/mnras/stw1433}, \href
  {https://ui.adsabs.harvard.edu/abs/2016MNRAS.461.2126S} {461, 2126}

\bibitem[\protect\citeauthoryear{{Sebesta}, {Williams}, {Liesenborgs},
  {Medezinski}  \& {Okabe}}{{Sebesta} et~al.}{2019}]{sebesta2019}
{Sebesta} K.,  {Williams} L. L.~R.,  {Liesenborgs} J.,  {Medezinski} E.,
  {Okabe} N.,  2019, \mn@doi [\mnras] {10.1093/mnras/stz1950}, \href
  {https://ui.adsabs.harvard.edu/abs/2019MNRAS.488.3251S} {488, 3251}

\bibitem[\protect\citeauthoryear{{Soucail}}{{Soucail}}{1987}]{Soucail1987}
{Soucail} G.,  1987, The Messenger, \href
  {https://ui.adsabs.harvard.edu/abs/1987Msngr..48...43S} {48, 43}

\bibitem[\protect\citeauthoryear{{Soucail}, {Mellier}, {Fort}, {Mathez}  \&
  {Cailloux}}{{Soucail} et~al.}{1988}]{Soucail1988}
{Soucail} G.,  {Mellier} Y.,  {Fort} B.,  {Mathez} G.,   {Cailloux} M.,  1988,
  \aap, \href {https://ui.adsabs.harvard.edu/abs/1988A&A...191L..19S} {191,
  L19}

\bibitem[\protect\citeauthoryear{{Steinhardt} et~al.,}{{Steinhardt}
  et~al.}{2020}]{Steinhardt2020}
{Steinhardt} C.~L.,  et~al., 2020, \mn@doi [\apjs] {10.3847/1538-4365/ab75ed},
  \href {https://ui.adsabs.harvard.edu/abs/2020ApJS..247...64S} {247, 64}

\bibitem[\protect\citeauthoryear{{Strait} et~al.,}{{Strait}
  et~al.}{2018}]{Strait2018}
{Strait} V.,  et~al., 2018, \mn@doi [\apj] {10.3847/1538-4357/aae834}, \href
  {https://ui.adsabs.harvard.edu/abs/2018ApJ...868..129S} {868, 129}

\bibitem[\protect\citeauthoryear{{Umetsu}, {Broadhurst}, {Zitrin}, {Medezinski}
   \& {Hsu}}{{Umetsu} et~al.}{2011}]{Umetsu2011}
{Umetsu} K.,  {Broadhurst} T.,  {Zitrin} A.,  {Medezinski} E.,   {Hsu} L.-Y.,
  2011, \mn@doi [\apj] {10.1088/0004-637X/729/2/127}, \href
  {https://ui.adsabs.harvard.edu/abs/2011ApJ...729..127U} {729, 127}

\bibitem[\protect\citeauthoryear{{Vega-Ferrero}, {Diego}  \&
  {Bernstein}}{{Vega-Ferrero} et~al.}{2019}]{Vega2019}
{Vega-Ferrero} J.,  {Diego} J.~M.,   {Bernstein} G.~M.,  2019, \mn@doi [\mnras]
  {10.1093/mnras/stz1217}, \href
  {https://ui.adsabs.harvard.edu/abs/2019MNRAS.486.5414V} {486, 5414}

\bibitem[\protect\citeauthoryear{{Williams} \& {Liesenborgs}}{{Williams} \&
  {Liesenborgs}}{2019}]{williams2019}
{Williams} L. L.~R.,  {Liesenborgs} J.,  2019, \mn@doi [\mnras]
  {10.1093/mnras/sty3113}, \href
  {https://ui.adsabs.harvard.edu/abs/2019MNRAS.482.5666W} {482, 5666}

\bibitem[\protect\citeauthoryear{{Williams}, {Sebesta}  \&
  {Liesenborgs}}{{Williams} et~al.}{2018}]{williams2018}
{Williams} L.~L.~R.,  {Sebesta} K.,   {Liesenborgs} J.,  2018, \mn@doi [\mnras]
  {10.1093/mnras/sty2088}, \href
  {https://ui.adsabs.harvard.edu/abs/2018MNRAS.480.3140W} {480, 3140}

\bibitem[\protect\citeauthoryear{{Wong}, {Ammons}, {Keeton}  \&
  {Zabludoff}}{{Wong} et~al.}{2012}]{Wong2012}
{Wong} K.~C.,  {Ammons} S.~M.,  {Keeton} C.~R.,   {Zabludoff} A.~I.,  2012,
  \mn@doi [\apj] {10.1088/0004-637X/752/2/104}, \href
  {https://ui.adsabs.harvard.edu/abs/2012ApJ...752..104W} {752, 104}

\bibitem[\protect\citeauthoryear{{Zitrin} \& {Broadhurst}}{{Zitrin} \&
  {Broadhurst}}{2016}]{Zitrin2016}
{Zitrin} A.,  {Broadhurst} T.,  2016, \mn@doi [\apj]
  {10.3847/0004-637X/833/1/25}, \href
  {https://ui.adsabs.harvard.edu/abs/2016ApJ...833...25Z} {833, 25}

\bibitem[\protect\citeauthoryear{{Zitrin} et~al.,}{{Zitrin}
  et~al.}{2009}]{Zitrin2009}
{Zitrin} A.,  et~al., 2009, \mn@doi [\mnras]
  {10.1111/j.1365-2966.2009.14899.x}, \href
  {https://ui.adsabs.harvard.edu/abs/2009MNRAS.396.1985Z} {396, 1985}

\bibitem[\protect\citeauthoryear{{Zitrin} et~al.,}{{Zitrin}
  et~al.}{2013}]{Zitrin2013}
{Zitrin} A.,  et~al., 2013, \mn@doi [\apjl] {10.1088/2041-8205/762/2/L30},
  \href {https://ui.adsabs.harvard.edu/abs/2013ApJ...762L..30Z} {762, L30}

\makeatother
\end{thebibliography}

%%%%%%%%%%%%%%%%%%%%%%%%%%%%%%%%%%%%%%%%%%%%%%%%%%

%%%%%%%%%%%%%%%%% APPENDICES %%%%%%%%%%%%%%%%%%%%%

\appendix

\section{Methods to compute the magnification distribution in the source plane}
\label{sec:magcomp}

\begin{figure}
	\includegraphics[width=\columnwidth]{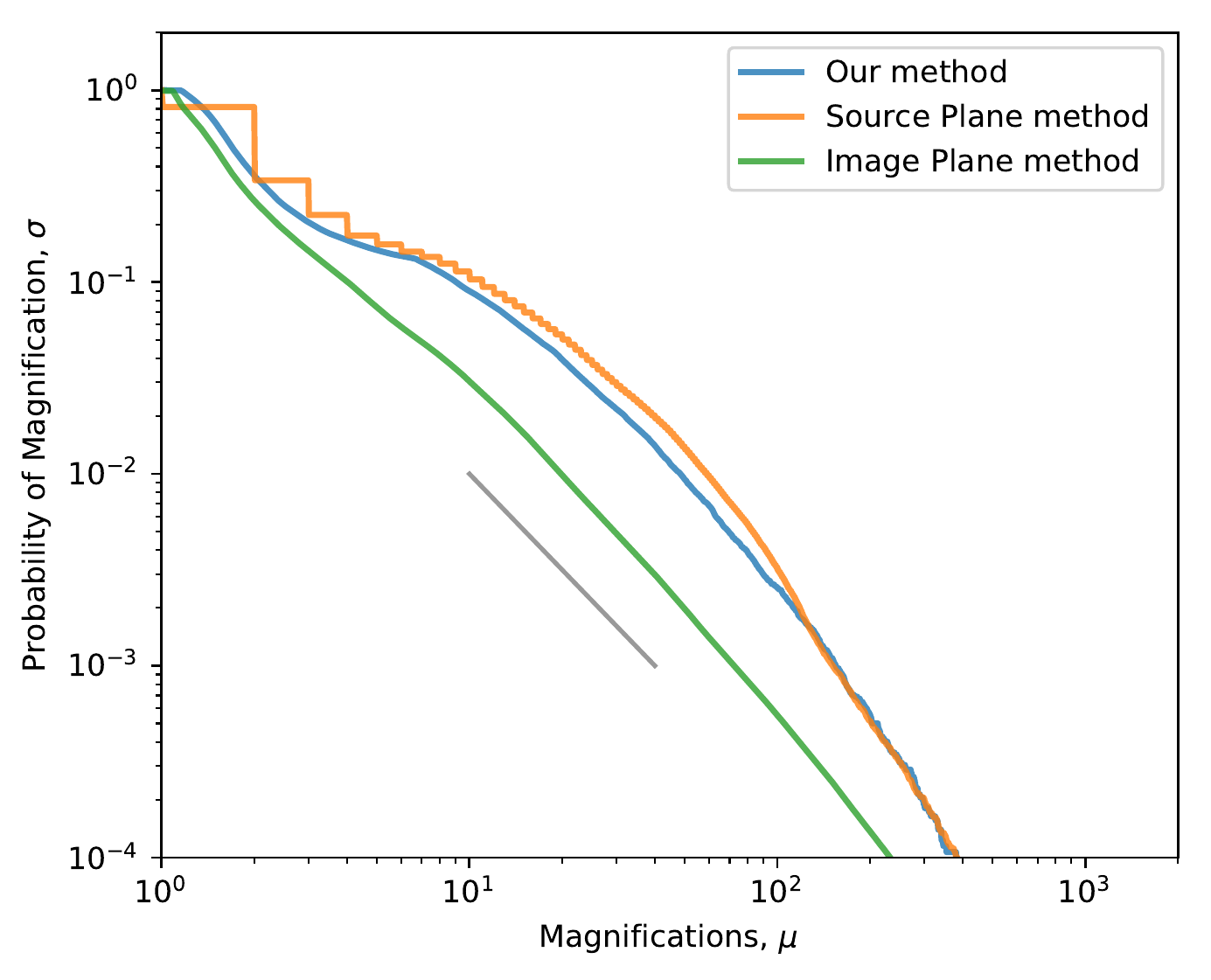}
    \caption{ { Comparison of the cumulative probability of magnifications in the source plane (similar to Figure~\ref{fig:mf}), as computed by our method and, the source plane method and the image plane method of \citet{Vega2019}, using the average \grale reconstruction of A370. The blue curves is the same as in Figure~\ref{fig:mf}. The gray line denotes a slope of $\mu^{-2}$.}}
    \label{fig:magcomp}
\end{figure}
Here we discuss a comparison between our method to compute the magnification distribution in the source plane and, the source plane method and the image plane method described in \cite{Vega2019}. 

In \cite{Vega2019}, the authors use the magnification distribution in the image plane to retrieve the magnification distribution in the source plane. This is done either by back-projecting the image plane pixels to the source plane (source plane method), or by computing the area of an source plane by diving the corresponding area of the image plane pixels by the magnification value $\mu$ associated with it (image plane method). Image plane method  is computationally cheaper of the two, and is widely used to compute magnification distribution in the source plane, for example, in the works of \citet{Johnson2014}, \citet{Richard2014}, \citet{Jauzac2015} et cetera. 

In contrast to both of these methods, we compute the magnification by forward lensing the source plane grid points using the reconstructed deflection angles (see Section~\ref{sec:magsp} for more details). Figure~\ref{fig:magcomp} shows a comparison of the results for cumulative probability of magnifications in the source plane as computed by our method, the source plane method and the image plane method, using the average \grale reconstruction of A370. As expected for a given mass distribution, our method and the source plane method produced very close results.

%%%%%%%%%%%%%%%%%%%%%%%%%%%%%%%%%%%%%%%%%%%%%%%%%%

% Don't change these lines
\bsp	% typesetting comment
\label{lastpage}
\end{document}